\documentclass[12pt,german]{article}
\usepackage[utf8]{inputenc}
\usepackage[T1]{fontenc}
\usepackage{lmodern}

\usepackage[margin=1.25in]{geometry}
\usepackage{amsmath,amssymb,amsthm,mathrsfs,mathtools}
\usepackage{bm}
\usepackage{setspace}
\usepackage{fancyhdr}
\usepackage{url}
\usepackage{color,graphicx}
\usepackage{hyperref}
\usepackage{qtree}
\usepackage{wrapfig}
\usepackage[linesnumbered,ruled,vlined]{algorithm2e}
\usepackage{comment}
\usepackage{etaremune}

\hypersetup{
    colorlinks,
    linktoc = page,
    linkcolor = {blue},
    citecolor = {blue},
    filecolor = {black},
    urlcolor = {cyan}
}
\usepackage{natbib}
\usepackage{tikz}
\usetikzlibrary{arrows,chains,positioning,scopes,calc,decorations.pathreplacing}
\tikzset{
  treenode/.style = {align=center, inner sep=0pt, text centered,
    font=\sffamily},
  arn_n/.style = {treenode, font=\sffamily\bfseries, draw=black,
    text width=1.5em},
  arn_r/.style = {treenode, circle, black, draw=black, 
    text width=1.5em},
  arn_y/.style = {treenode, circle,  black, draw=black, text width=1.5em, fill = block-gray}
}
\usepackage[most]{tcolorbox}
\definecolor{block-gray}{gray}{0.85}
\newtcolorbox{blockquote}{colback=block-gray,grow to right by=-1mm,grow to left by=-1mm,boxrule=0pt,boxsep=0pt,breakable}

\newtheorem{theorem}{Theorem}
\newtheorem{assumption}{Assumption}

\theoremstyle{remark}

\title{Inference for microbe--metabolite association networks using a latent graph model}

\author{Jing Ma\footnote{Correspondence: {jingma@fredhutch.org}}
\\
Division of Public Health Sciences, Fred Hutchinson Cancer Center }

\begin{document}

\newcommand{\comments}[1]{}
\newcommand{\itemph}[1]{{\it\emph{#1}}}

%
%
%
%
%
\def\bzero{{\bf 0}}
\def\bone{{\bf 1}}

%
%
\def\blue{\color{blue}}
\def\red{\color{red}}
\def\rank{{\rm rank}}
\def\eqtl{{\rm eQTL}}
\def\nb{{\rm ne}}
\def\clr{{\rm clr}}
\def\rclr{{\rm rclr}}
\def\alr{{\rm alr}}
\def\ilr{{\rm ilr}}
\def\LNPN{{\rm LNPN}}
\def\NPN{{\rm NPN}}
\def\close{\mathcal{C}}
\def\supp{{\rm supp}}
\def\sign{\mbox{sign}}
%
%
\def\ba{{\mbox{\boldmath$a$}}}
\def\bb{{\boldsymbol b}}
\def\bc{{\boldsymbol c}}
\def\bd{{\boldsymbol d}}
\def\be{{\boldsymbol e}}
\def\bdf{{\boldsymbol f}}
\def\bg{{\mbox{\boldmath$g$}}}
\def\bh{{\boldsymbol h}}
\def\bi{{\boldsymbol i}}
\def\bj{{\boldsymbol j}}
\def\bk{{\boldsymbol k}}
\def\bl{{\boldsymbol l}}
\def\bm{{\boldsymbol m}}
\def\bn{{\boldsymbol n}}
\def\bo{{\boldsymbol o}}
\def\bp{{\boldsymbol p}}
\def\bq{{\boldsymbol q}}
\def\br{{\boldsymbol r}}
\def\bs{{\boldsymbol s}}
\def\bt{{\boldsymbol t}}
\def\bu{{\boldsymbol u}}
\def\bv{{\boldsymbol v}}
\def\bw{{\boldsymbol w}}
\def\bx{{\boldsymbol x}}
\def\by{{\boldsymbol y}}
\def\bz{{\boldsymbol z}}
\def\bA{{\bf A}}
\def\bB{{\bf B}}
\def\bC{{\bf C}}
\def\bD{{\bf D}}
\def\bE{{\bf E}}
\def\bF{{\bf F}}
\def\bG{{\bf G}}
\def\bH{{\bf H}}
\def\bI{{\bf I}}
\def\bJ{{\bf J}}
\def\bK{{\bf K}}
\def\bL{{\bf L}}
\def\bM{{\bf M}}
\def\bN{{\bf N}}
\def\bO{{\bf O}}
\def\bP{{\bf P}}
\def\bQ{{\bf Q}}
\def\bR{{\bf R}}
\def\bS{{\bf S}}
\def\bT{{\bf T}}
\def\bU{{\bf U}}
\def\bV{{\bf V}}
\def\bW{{\bf W}}
\def\bX{{\bf X}}
\def\bY{{\bf Y}}
\def\bZ{{\bf Z}}
\def\smbZ{\scriptstyle{\bf Z}}
\def\smM{\scriptstyle{M}}
\def\smN{\scriptstyle{N}}
\def\smbT{\scriptstyle{\bf T}}
%
%
%
%
\def\eps{\varepsilon}
\def\thick#1{\hbox{\rlap{$#1$}\kern0.25pt\rlap{$#1$}\kern0.25pt$#1$}}
\def\balpha{\boldsymbol{\alpha}}
\def\bbeta{\boldsymbol{\beta}}
\def\bgamma{\boldsymbol{\gamma}}
\def\bdelta{\boldsymbol{\delta}}
\def\btheta{\boldsymbol{\theta}}
\def\bepsilon{\boldsymbol{\epsilon}}
\def\bvarepsilon{\boldsymbol{\varepsilon}}
\def\bzeta{\boldsymbol{\zeta}}
\def\bdeta{\boldsymbol{\eta}}
\def\biota{\boldsymbol{\iota}}
\def\bkappa{\boldsymbol{\kappa}}
\def\blambda{\boldsymbol{\lambda}}
\def\bmu{\boldsymbol{\mu}}
\def\bnu{\boldsymbol{\nu}}
\def\bxi{\boldsymbol{\xi}}
\def\bomicron{\boldsymbol{\omicron}}
\def\bpi{\boldsymbol{\pi}}
\def\brho{\boldsymbol{\rho}}
\def\bsigma{\boldsymbol{\sigma}}
\def\btau{\boldsymbol{\tau}}
\def\bupsilon{\boldsymbol{\upsilon}}
\def\bphi{\boldsymbol{\phi}}
\def\bchi{\boldsymbol{\chi}}
\def\bpsi{\boldsymbol{\psi}}
\def\bomega{\boldsymbol{\omega}}
\def\bAlpha{\boldsymbol{\Alpha}}
\def\bBeta{\boldsymbol{\Beta}}
\def\bGamma{\boldsymbol{\Gamma}}
\def\bDelta{\boldsymbol{\Delta}}
\def\bEpsilon{\boldsymbol{\Epsilon}}
\def\bZeta{\boldsymbol{\Zeta}}
\def\bEta{\boldsymbol{\Eta}}
\def\bTheta{\boldsymbol{\Theta}}
\def\bIota{\boldsymbol{\Iota}}
\def\bKappa{\boldsymbol{\Kappa}}
\def\bLambda{{\boldsymbol{\Lambda}}}
\def\bMu{\boldsymbol{\Mu}}
\def\bNu{\boldsymbol{\Nu}}
\def\bXi{\boldsymbol{\Xi}}
\def\bOmicron{\boldsymbol{\Omicron}}
\def\bPi{\boldsymbol{\Pi}}
\def\bRho{\boldsymbol{\Rho}}
\def\bSigma{\boldsymbol{\Sigma}}
\def\bTau{\boldsymbol{\Tau}}
\def\bUpsilon{\boldsymbol{\Upsilon}}
\def\bPhi{\boldsymbol{\Phi}}
\def\bChi{\boldsymbol{\Chi}}
\def\bPsi{\boldsymbol{\Psi}}
\def\bOmega{\boldsymbol{\Omega}}
%
%
%
\def\smalpha{{{\scriptstyle{\alpha}}}}
\def\smbeta{{{\scriptstyle{\beta}}}}
\def\smgamma{{{\scriptstyle{\gamma}}}}
\def\smdelta{{{\scriptstyle{\delta}}}}
\def\smepsilon{{{\scriptstyle{\epsilon}}}}
\def\smvarepsilon{{{\scriptstyle{\varepsilon}}}}
\def\smzeta{{{\scriptstyle{\zeta}}}}
\def\smdeta{{{\scriptstyle{\eta}}}}
\def\smtheta{{{\scriptstyle{\theta}}}}
\def\smiota{{{\scriptstyle{\iota}}}}
\def\smkappa{{{\scriptstyle{\kappa}}}}
\def\smlambda{{{\scriptstyle{\lambda}}}}
\def\smmu{{{\scriptstyle{\mu}}}}
\def\smnu{{{\scriptstyle{\nu}}}}
\def\smxi{{{\scriptstyle{\xi}}}}
\def\smomicron{{{\scriptstyle{\omicron}}}}
\def\smpi{{{\scriptstyle{\pi}}}}
\def\smrho{{{\scriptstyle{\rho}}}}
\def\smsigma{{{\scriptstyle{\sigma}}}}
\def\smtau{{{\scriptstyle{\tau}}}}
\def\smupsilon{{{\scriptstyle{\upsilon}}}}
\def\smphi{{{\scriptstyle{\phi}}}}
\def\smchi{{{\scriptstyle{\chi}}}}
\def\smpsi{{{\scriptstyle{\psi}}}}
\def\smomega{{{\scriptstyle{\omega}}}}
\def\smAlpha{{{\scriptstyle{\Alpha}}}}
\def\smBeta{{{\scriptstyle{\Beta}}}}
\def\smGamma{{{\scriptstyle{\Gamma}}}}
\def\smDelta{{{\scriptstyle{\Delta}}}}
\def\smEpsilon{{{\scriptstyle{\Epsilon}}}}
\def\smZeta{{{\scriptstyle{\Zeta}}}}
\def\smEta{{{\scriptstyle{\Eta}}}}
\def\smTheta{{{\scriptstyle{\Theta}}}}
\def\smIota{{{\scriptstyle{\Iota}}}}
\def\smKappa{{{\scriptstyle{\Kappa}}}}
\def\smLambda{{{\scriptstyle{\Lambda}}}}
\def\smMu{{{\scriptstyle{\Mu}}}}
\def\smNu{{{\scriptstyle{\Nu}}}}
\def\smXi{{{\scriptstyle{\Xi}}}}
\def\smOmicron{{{\scriptstyle{\Omicron}}}}
\def\smPi{{{\scriptstyle{\Pi}}}}
\def\smRho{{{\scriptstyle{\Rho}}}}
\def\smSigma{{{\scriptstyle{\Sigma}}}}
\def\smTau{{{\scriptstyle{\Tau}}}}
\def\smUpsilon{{{\scriptstyle{\Upsilon}}}}
\def\smPhi{{{\scriptstyle{\Phi}}}}
\def\smChi{{{\scriptstyle{\Chi}}}}
\def\smPsi{{{\scriptstyle{\Psi}}}}
\def\smOmega{{{\scriptstyle{\Omega}}}}
%
%

%
\def\smbalpha{\boldsymbol{{\scriptstyle{\alpha}}}}
\def\smbbeta{\boldsymbol{{\scriptstyle{\beta}}}}
\def\smbgamma{\boldsymbol{{\scriptstyle{\gamma}}}}
\def\smbdelta{\boldsymbol{{\scriptstyle{\delta}}}}
\def\smbepsilon{\boldsymbol{{\scriptstyle{\epsilon}}}}
\def\smbvarepsilon{\boldsymbol{{\scriptstyle{\varepsilon}}}}
\def\smbzeta{\boldsymbol{{\scriptstyle{\zeta}}}}
\def\smbdeta{\boldsymbol{{\scriptstyle{\eta}}}}
\def\smbtheta{\boldsymbol{{\scriptstyle{\theta}}}}
\def\smbiota{\boldsymbol{{\scriptstyle{\iota}}}}
\def\smbkappa{\boldsymbol{{\scriptstyle{\kappa}}}}
\def\smblambda{\boldsymbol{{\scriptstyle{\lambda}}}}
\def\smbmu{\boldsymbol{{\scriptstyle{\mu}}}}
\def\smbnu{\boldsymbol{{\scriptstyle{\nu}}}}
\def\smbxi{\boldsymbol{{\scriptstyle{\xi}}}}
\def\smbomicron{\boldsymbol{{\scriptstyle{\omicron}}}}
\def\smbpi{\boldsymbol{{\scriptstyle{\pi}}}}
\def\smbrho{\boldsymbol{{\scriptstyle{\rho}}}}
\def\smbsigma{\boldsymbol{{\scriptstyle{\sigma}}}}
\def\smbtau{\boldsymbol{{\scriptstyle{\tau}}}}
\def\smbupsilon{\boldsymbol{{\scriptstyle{\upsilon}}}}
\def\smbphi{\boldsymbol{{\scriptstyle{\phi}}}}
\def\smbchi{\boldsymbol{{\scriptstyle{\chi}}}}
\def\smbpsi{\boldsymbol{{\scriptstyle{\psi}}}}
\def\smbomega{\boldsymbol{{\scriptstyle{\omega}}}}
\def\smbAlpha{\boldsymbol{{\scriptstyle{\Alpha}}}}
\def\smbBeta{\boldsymbol{{\scriptstyle{\Beta}}}}
\def\smbGamma{\boldsymbol{{\scriptstyle{\Gamma}}}}
\def\smbDelta{\boldsymbol{{\scriptstyle{\Delta}}}}
\def\smbEpsilon{\boldsymbol{{\scriptstyle{\Epsilon}}}}
\def\smbZeta{\boldsymbol{{\scriptstyle{\Zeta}}}}
\def\smbEta{\boldsymbol{{\scriptstyle{\Eta}}}}
\def\smbTheta{\boldsymbol{{\scriptstyle{\Theta}}}}
\def\smbIota{\boldsymbol{{\scriptstyle{\Iota}}}}
\def\smbKappa{\boldsymbol{{\scriptstyle{\Kappa}}}}
\def\smbLambda{\boldsymbol{{\scriptstyle{\Lambda}}}}
\def\smbMu{\boldsymbol{{\scriptstyle{\Mu}}}}
\def\smbNu{\boldsymbol{{\scriptstyle{\Nu}}}}
\def\smbXi{\boldsymbol{{\scriptstyle{\Xi}}}}
\def\smbOmicron{\boldsymbol{{\scriptstyle{\Omicron}}}}
\def\smbPi{\boldsymbol{{\scriptstyle{\Pi}}}}
\def\smbRho{\boldsymbol{{\scriptstyle{\Rho}}}}
\def\smbSigma{\boldsymbol{{\scriptstyle{\Sigma}}}}
\def\smbTau{\boldsymbol{{\scriptstyle{\Tau}}}}
\def\smbUpsilon{\boldsymbol{{\scriptstyle{\Upsilon}}}}
\def\smbPhi{\boldsymbol{{\scriptstyle{\Phi}}}}
\def\smbChi{\boldsymbol{{\scriptstyle{\Chi}}}}
\def\smbPsi{\boldsymbol{{\scriptstyle{\Psi}}}}
\def\smbOmega{\boldsymbol{{\scriptstyle{\Omega}}}}
%
%
%
%
\def\ahat{{\widehat a}}
\def\bhat{{\widehat b}}
\def\chat{{\widehat c}}
\def\dhat{{\widehat d}}
\def\ehat{{\widehat e}}
\def\fhat{{\widehat f}}
\def\ghat{{\widehat g}}
\def\hhat{{\widehat h}}
\def\ihat{{\widehat i}}
\def\jhat{{\widehat j}}
\def\khat{{\widehat k}}
\def\lhat{{\widehat l}}
\def\mhat{{\widehat m}}
\def\nhat{{\widehat n}}
\def\ohat{{\widehat o}}
\def\phat{{\widehat p}}
\def\qhat{{\widehat q}}
\def\rhat{{\widehat r}}
\def\shat{{\widehat s}}
\def\that{{\widehat t}}
\def\uhat{{\widehat u}}
\def\vhat{{\widehat v}}
\def\what{{\widehat w}}
\def\xhat{{\widehat x}}
\def\yhat{{\widehat y}}
\def\zhat{{\widehat z}}
\def\Ahat{{\widehat A}}
\def\Bhat{{\widehat B}}
\def\Chat{{\widehat C}}
\def\Dhat{{\widehat D}}
\def\Ehat{{\widehat E}}
\def\Fhat{{\widehat F}}
\def\Ghat{{\widehat G}}
\def\Hhat{{\widehat H}}
\def\Ihat{{\widehat I}}
\def\Jhat{{\widehat J}}
\def\Khat{{\widehat K}}
\def\Lhat{{\widehat L}}
\def\Mhat{{\widehat M}}
\def\Nhat{{\widehat N}}
\def\Ohat{{\widehat O}}
\def\Phat{{\widehat P}}
\def\Qhat{{\widehat Q}}
\def\Rhat{{\widehat R}}
\def\Shat{{\widehat S}}
\def\That{{\widehat T}}
\def\Uhat{{\widehat U}}
\def\Vhat{{\widehat V}}
\def\What{{\widehat W}}
\def\Xhat{{\widehat X}}
\def\Yhat{{\widehat Y}}
\def\Zhat{{\widehat Z}}
%
%
%
\def\atilde{{\widetilde a}}
\def\btilde{{\widetilde b}}
\def\ctilde{{\widetilde c}}
\def\dtilde{{\widetilde d}}
\def\etilde{{\widetilde e}}
\def\ftilde{{\widetilde f}}
\def\gtilde{{\widetilde g}}
\def\htilde{{\widetilde h}}
\def\itilde{{\widetilde i}}
\def\jtilde{{\widetilde j}}
\def\ktilde{{\widetilde k}}
\def\ltilde{{\widetilde l}}
\def\mtilde{{\widetilde m}}
\def\ntilde{{\widetilde n}}
\def\otilde{{\widetilde o}}
\def\ptilde{{\widetilde p}}
\def\qtilde{{\widetilde q}}
\def\rtilde{{\widetilde r}}
\def\stilde{{\widetilde s}}
\def\ttilde{{\widetilde t}}
\def\utilde{{\widetilde u}}
\def\vtilde{{\widetilde v}}
\def\wtilde{{\widetilde w}}
\def\xtilde{{\widetilde x}}
\def\ytilde{{\widetilde y}}
\def\ztilde{{\widetilde z}}
\def\Atilde{{\widetilde A}}
\def\Btilde{{\widetilde B}}
\def\Ctilde{{\widetilde C}}
\def\Dtilde{{\widetilde D}}
\def\Etilde{{\widetilde E}}
\def\Ftilde{{\widetilde F}}
\def\Gtilde{{\widetilde G}}
\def\Htilde{{\widetilde H}}
\def\Itilde{{\widetilde I}}
\def\Jtilde{{\widetilde J}}
\def\Ktilde{{\widetilde K}}
\def\Ltilde{{\widetilde L}}
\def\Mtilde{{\widetilde M}}
\def\Ntilde{{\widetilde N}}
\def\Otilde{{\widetilde O}}
\def\Ptilde{{\widetilde P}}
\def\Qtilde{{\widetilde Q}}
\def\Rtilde{{\widetilde R}}
\def\Stilde{{\widetilde S}}
\def\Ttilde{{\widetilde T}}
\def\Utilde{{\widetilde U}}
\def\Vtilde{{\widetilde V}}
\def\Wtilde{{\widetilde W}}
\def\Xtilde{{\widetilde X}}
\def\Ytilde{{\widetilde Y}}
\def\Ztilde{{\widetilde Z}}
%
%
%
%
\def\bahat{{\widehat \ba}}
\def\bbhat{{\widehat \bb}}
\def\bchat{{\widehat \bc}}
\def\bdhat{{\widehat \bd}}
\def\behat{{\widehat \be}}
\def\bfhat{{\widehat \bf}}
\def\bghat{{\widehat \bg}}
\def\bhhat{{\widehat \bh}}
\def\bihat{{\widehat \bi}}
\def\bjhat{{\widehat \bj}}
\def\bkhat{{\widehat \bk}}
\def\blhat{{\widehat \bl}}
\def\bmhat{{\widehat \bm}}
\def\bnhat{{\widehat \bn}}
\def\bohat{{\widehat \bo}}
\def\bphat{{\widehat \bp}}
\def\bqhat{{\widehat \bq}}
\def\brhat{{\widehat \br}}
\def\bshat{{\widehat \bs}}
\def\bthat{{\widehat \bt}}
\def\buhat{{\widehat \bu}}
\def\bvhat{{\widehat \bv}}
\def\bwhat{{\widehat \bw}}
\def\bxhat{{\widehat \bx}}
\def\byhat{{\widehat \by}}
\def\bzhat{{\widehat \bz}}
\def\bAhat{{\widehat \bA}}
\def\bBhat{{\widehat \bB}}
\def\bChat{{\widehat \bC}}
\def\bDhat{{\widehat \bD}}
\def\bEhat{{\widehat \bE}}
\def\bFhat{{\widehat \bF}}
\def\bGhat{{\widehat \bG}}
\def\bHhat{{\widehat \bH}}
\def\bIhat{{\widehat \bI}}
\def\bJhat{{\widehat \bJ}}
\def\bKhat{{\widehat \bK}}
\def\bLhat{{\widehat \bL}}
\def\bMhat{{\widehat \bM}}
\def\bNhat{{\widehat \bN}}
\def\bOhat{{\widehat \bO}}
\def\bPhat{{\widehat \bP}}
\def\bQhat{{\widehat \bQ}}
\def\bRhat{{\widehat \bR}}
\def\bShat{{\widehat \bS}}
\def\bThat{{\widehat \bT}}
\def\bUhat{{\widehat \bU}}
\def\bVhat{{\widehat \bV}}
\def\bWhat{{\widehat \bW}}
\def\bXhat{{\widehat \bX}}
\def\bYhat{{\widehat \bY}}
\def\bZhat{{\widehat \bZ}}
%
%
%
%
%
\def\batilde{{\widetilde \ba}}
\def\bbtilde{{\widetilde \bb}}
\def\bctilde{{\widetilde \bc}}
\def\bdtilde{{\widetilde \bd}}
\def\betilde{{\widetilde \be}}
\def\bftilde{{\widetilde \bf}}
\def\bgtilde{{\widetilde \bg}}
\def\bhtilde{{\widetilde \bh}}
\def\bitilde{{\widetilde \bi}}
\def\bjtilde{{\widetilde \bj}}
\def\bktilde{{\widetilde \bk}}
\def\bltilde{{\widetilde \bl}}
\def\bmtilde{{\widetilde \bm}}
\def\bntilde{{\widetilde \bn}}
\def\botilde{{\widetilde \bo}}
\def\bptilde{{\widetilde \bp}}
\def\bqtilde{{\widetilde \bq}}
\def\brtilde{{\widetilde \br}}
\def\bstilde{{\widetilde \bs}}
\def\bttilde{{\widetilde \bt}}
\def\butilde{{\widetilde \bu}}
\def\bvtilde{{\widetilde \bv}}
\def\bwtilde{{\widetilde \bw}}
\def\bxtilde{{\widetilde \bx}}
\def\bytilde{{\widetilde \by}}
\def\bztilde{{\widetilde \bz}}
\def\bAtilde{{\widetilde \bA}}
\def\bBtilde{{\widetilde \bB}}
\def\bCtilde{{\widetilde \bC}}
\def\bDtilde{{\widetilde \bD}}
\def\bEtilde{{\widetilde \bE}}
\def\bFtilde{{\widetilde \bF}}
\def\bGtilde{{\widetilde \bG}}
\def\bHtilde{{\widetilde \bH}}
\def\bItilde{{\widetilde \bI}}
\def\bJtilde{{\widetilde \bJ}}
\def\bKtilde{{\widetilde \bK}}
\def\bLtilde{{\widetilde \bL}}
\def\bMtilde{{\widetilde \bM}}
\def\bNtilde{{\widetilde \bN}}
\def\bOtilde{{\widetilde \bO}}
\def\bPtilde{{\widetilde \bP}}
\def\bQtilde{{\widetilde \bQ}}
\def\bRtilde{{\widetilde \bR}}
\def\bStilde{{\widetilde \bS}}
\def\bTtilde{{\widetilde \bT}}
\def\bUtilde{{\widetilde \bU}}
\def\bVtilde{{\widetilde \bV}}
\def\bWtilde{{\widetilde \bW}}
\def\bXtilde{{\widetilde \bX}}
\def\bYtilde{{\widetilde \bY}}
\def\bZtilde{{\widetilde \bZ}}
%
%
%
%
%
%
\def\alphahat{{\widehat\alpha}}
\def\betahat{{\widehat\beta}}
\def\gammahat{{\widehat\gamma}}
\def\deltahat{{\widehat\delta}}
\def\epsilonhat{{\widehat\epsilon}}
\def\varepsilonhat{{\widehat\varepsilon}}
\def\zetahat{{\widehat\zeta}}
\def\etahat{{\widehat\eta}}
\def\thetahat{{\widehat\theta}}
\def\iotahat{{\widehat\iota}}
\def\kappahat{{\widehat\kappa}}
\def\lambdahat{{\widehat\lambda}}
\def\muhat{{\widehat\mu}}
\def\nuhat{{\widehat\nu}}
\def\xihat{{\widehat\xi}}
\def\omicronhat{{\widehat\omicron}}
\def\pihat{{\widehat\pi}}
\def\rhohat{{\widehat\rho}}
\def\sigmahat{{\widehat\sigma}}
\def\tauhat{{\widehat\tau}}
\def\upsilonhat{{\widehat\upsilon}}
\def\phihat{{\widehat\phi}}
\def\chihat{{\widehat\chi}}
\def\psihat{{\widehat\psi}}
\def\omegahat{{\widehat\omega}}
\def\Alphahat{{\widehat\Alpha}}
\def\Betahat{{\widehat\Beta}}
\def\Gammahat{{\widehat\Gamma}}
\def\Deltahat{{\widehat\Delta}}
\def\Epsilonhat{{\widehat\Epsilon}}
\def\Zetahat{{\widehat\Zeta}}
\def\Etahat{{\widehat\Eta}}
\def\Thetahat{{\widehat\Theta}}
\def\Iotahat{{\widehat\Iota}}
\def\Kappahat{{\widehat\Kappa}}
\def\Lambdahat{{\widehat\Lambda}}
\def\Muhat{{\widehat\Mu}}
\def\Nuhat{{\widehat\Nu}}
\def\Xihat{{\widehat\Xi}}
\def\Omicronhat{{\widehat\Omicron}}
\def\Pihat{{\widehat\Pi}}
\def\Rhohat{{\widehat\Rho}}
\def\Sigmahat{{\widehat\Sigma}}
\def\Tauhat{{\widehat\Tau}}
\def\Upsilonhat{{\widehat\Upsilon}}
\def\Phihat{{\widehat\Phi}}
\def\Chihat{{\widehat\Chi}}
\def\Psihat{{\widehat\Psi}}
\def\Omegahat{{\widehat\Omega}}
%
%
%
%
%
\def\alphatilde{{\widetilde\alpha}}
\def\betatilde{{\widetilde\beta}}
\def\gammatilde{{\widetilde\gamma}}
\def\deltatilde{{\widetilde\delta}}
\def\epsilontilde{{\widetilde\epsilon}}
\def\varepsilontilde{{\widetilde\varepsilon}}
\def\zetatilde{{\widetilde\zeta}}
\def\etatilde{{\widetilde\eta}}
\def\thetatilde{{\widetilde\theta}}
\def\iotatilde{{\widetilde\iota}}
\def\kappatilde{{\widetilde\kappa}}
\def\lambdatilde{{\widetilde\lambda}}
\def\mutilde{{\widetilde\mu}}
\def\nutilde{{\widetilde\nu}}
\def\xitilde{{\widetilde\xi}}
\def\omicrontilde{{\widetilde\omicron}}
\def\pitilde{{\widetilde\pi}}
\def\rhotilde{{\widetilde\rho}}
\def\sigmatilde{{\widetilde\sigma}}
\def\tautilde{{\widetilde\tau}}
\def\upsilontilde{{\widetilde\upsilon}}
\def\phitilde{{\widetilde\phi}}
\def\chitilde{{\widetilde\chi}}
\def\psitilde{{\widetilde\psi}}
\def\omegatilde{{\widetilde\omega}}
\def\Alphatilde{{\widetilde\Alpha}}
\def\Betatilde{{\widetilde\Beta}}
\def\Gammatilde{{\widetilde\Gamma}}
\def\Deltatilde{{\widetilde\Delta}}
\def\Epsilontilde{{\widetilde\Epsilon}}
\def\Zetatilde{{\widetilde\Zeta}}
\def\Etatilde{{\widetilde\Eta}}
\def\Thetatilde{{\widetilde\Theta}}
\def\Iotatilde{{\widetilde\Iota}}
\def\Kappatilde{{\widetilde\Kappa}}
\def\Lambdatilde{{\widetilde\Lambda}}
\def\Mutilde{{\widetilde\Mu}}
\def\Nutilde{{\widetilde\Nu}}
\def\Xitilde{{\widetilde\Xi}}
\def\Omicrontilde{{\widetilde\Omicron}}
\def\Pitilde{{\widetilde\Pi}}
\def\Rhotilde{{\widetilde\Rho}}
\def\Sigmatilde{{\widetilde\Sigma}}
\def\Tautilde{{\widetilde\Tau}}
\def\Upsilontilde{{\widetilde\Upsilon}}
\def\Phitilde{{\widetilde\Phi}}
\def\Chitilde{{\widetilde\Chi}}
\def\Psitilde{{\widetilde\Psi}}
\def\Omegatilde{{\widetilde\Omega}}
%
%
%
%
%
%
\def\balphahat{{\widehat\balpha}}
\def\bbetahat{{\widehat\bbeta}}
\def\bgammahat{{\widehat\bgamma}}
\def\bdeltahat{{\widehat\bdelta}}
\def\bepsilonhat{{\widehat\bepsilon}}
\def\bzetahat{{\widehat\bzeta}}
\def\bdetahat{{\widehat\bdeta}}
\def\bthetahat{{\widehat\btheta}}
\def\biotahat{{\widehat\biota}}
\def\bkappahat{{\widehat\bkappa}}
\def\blambdahat{{\widehat\blambda}}
\def\bmuhat{{\widehat\bmu}}
\def\bnuhat{{\widehat\bnu}}
\def\bxihat{{\widehat\bxi}}
\def\bomicronhat{{\widehat\bomicron}}
\def\bpihat{{\widehat\bpi}}
\def\brhohat{{\widehat\brho}}
\def\bsigmahat{{\widehat\bsigma}}
\def\btauhat{{\widehat\btau}}
\def\bupsilonhat{{\widehat\bupsilon}}
\def\bphihat{{\widehat\bphi}}
\def\bchihat{{\widehat\bchi}}
\def\bpsihat{{\widehat\bpsi}}
\def\bomegahat{{\widehat\bomega}}
\def\bAlphahat{{\widehat\bAlpha}}
\def\bBetahat{{\widehat\bBeta}}
\def\bGammahat{{\widehat\bGamma}}
\def\bDeltahat{{\widehat\bDelta}}
\def\bEpsilonhat{{\widehat\bEpsilon}}
\def\bZetahat{{\widehat\bZeta}}
\def\bEtahat{{\widehat\bEta}}
\def\bThetahat{{\widehat\bTheta}}
\def\bIotahat{{\widehat\bIota}}
\def\bKappahat{{\widehat\bKappa}}
\def\bLambdahat{{\widehat\bLambda}}
\def\bMuhat{{\widehat\bMu}}
\def\bNuhat{{\widehat\bNu}}
\def\bXihat{{\widehat\bXi}}
\def\bOmicronhat{{\widehat\bOmicron}}
\def\bPihat{{\widehat\bPi}}
\def\bRhohat{{\widehat\bRho}}
\def\bSigmahat{{\widehat\bSigma}}
\def\bTauhat{{\widehat\bTau}}
\def\bUpsilonhat{{\widehat\bUpsilon}}
\def\bPhihat{{\widehat\bPhi}}
\def\bChihat{{\widehat\bChi}}
\def\bPsihat{{\widehat\bPsi}}
\def\bOmegahat{{\widehat\bOmega}}%
\def\balphahattrans{{\balphahat^{_{\transpose}}}}
\def\bbetahattrans{{\bbetahat^{_{\transpose}}}}
\def\bgammahattrans{{\bgammahat^{_{\transpose}}}}
\def\bdeltahattrans{{\bdeltahat^{_{\transpose}}}}
\def\bepsilonhattrans{{\bepsilonhat^{_{\transpose}}}}
\def\bzetahattrans{{\bzetahat^{_{\transpose}}}}
\def\bdetahattrans{{\bdetahat^{_{\transpose}}}}
\def\bthetahattrans{{\bthetahat^{_{\transpose}}}}
\def\biotahattrans{{\biotahat^{_{\transpose}}}}
\def\bkappahattrans{{\bkappahat^{_{\transpose}}}}
\def\blambdahattrans{{\blambdahat^{_{\transpose}}}}
\def\bmuhattrans{{\bmuhat^{_{\transpose}}}}
\def\bnuhattrans{{\bnuhat^{_{\transpose}}}}
\def\bxihattrans{{\bxihat^{_{\transpose}}}}
\def\bomicronhattrans{{\bomicronhat^{_{\transpose}}}}
\def\bpihattrans{{\bpihat^{_{\transpose}}}}
\def\brhohattrans{{\brhohat^{_{\transpose}}}}
\def\bsigmahattrans{{\bsigmahat^{_{\transpose}}}}
\def\btauhattrans{{\btauhat^{_{\transpose}}}}
\def\bupsilonhattrans{{\bupsilonhat^{_{\transpose}}}}
\def\bphihattrans{{\bphihat^{_{\transpose}}}}
\def\bchihattrans{{\bchihat^{_{\transpose}}}}
\def\bpsihattrans{{\bpsihat^{_{\transpose}}}}
\def\bomegahattrans{{\bomegahat^{_{\transpose}}}}
\def\bAlphahattrans{{\bAlphahat^{_{\transpose}}}}
\def\bBetahattrans{{\bBetahat^{_{\transpose}}}}
\def\bGammahattrans{{\bGammahat^{_{\transpose}}}}
\def\bDeltahattrans{{\bDeltahat^{_{\transpose}}}}
\def\bEpsilonhattrans{{\bEpsilonhat^{_{\transpose}}}}
\def\bZetahattrans{{\bZetahat^{_{\transpose}}}}
\def\bEtahattrans{{\bEtahat^{_{\transpose}}}}
\def\bThetahattrans{{\bThetahat^{_{\transpose}}}}
\def\bIotahattrans{{\bIotahat^{_{\transpose}}}}
\def\bKappahattrans{{\bKappahat^{_{\transpose}}}}
\def\bLambdahattrans{{\bLambdahat^{_{\transpose}}}}
\def\bMuhattrans{{\bMuhat^{_{\transpose}}}}
\def\bNuhattrans{{\bNuhat^{_{\transpose}}}}
\def\bXihattrans{{\bXihat^{_{\transpose}}}}
\def\bOmicronhattrans{{\bOmicronhat^{_{\transpose}}}}
\def\bPihattrans{{\bPihat^{_{\transpose}}}}
\def\bRhohattrans{{\bRhohat^{_{\transpose}}}}
\def\bSigmahattrans{{\bSigmahat^{_{\transpose}}}}
\def\bTauhattrans{{\bTauhat^{_{\transpose}}}}
\def\bUpsilonhattrans{{\bUpsilonhat^{_{\transpose}}}}
\def\bPhihattrans{{\bPhihat^{_{\transpose}}}}
\def\bChihattrans{{\bChihat^{_{\transpose}}}}
\def\bPsihattrans{{\bPsihat^{_{\transpose}}}}
\def\bOmegahattrans{{\bOmegahat^{_{\transpose}}}}%
%
\def\smbalpha{\widehat{\smbalpha}}
\def\smbbetahat{\widehat{\smbbeta}}
\def\smbgammahat{\widehat{\smbgamma}}
\def\smbdeltahat{\widehat{\smbdelta}}
\def\smbepsilonhat{\widehat{\smbepsilon}}
\def\smbvarepsilonhat{\widehat{\smbvarepsilon}}
\def\smbzetahat{\widehat{\smbzeta}}
\def\smbdetahat{\widehat{\smbeta}}
\def\smbthetahat{\widehat{\smbtheta}}
\def\smbiotahat{\widehat{\smbiota}}
\def\smbkappahat{\widehat{\smbkappa}}
\def\smblambdahat{\widehat{\smblambda}}
\def\smbmuhat{\widehat{\smbmu}}
\def\smbnuhat{\widehat{\smbnu}}
\def\smbxihat{\widehat{\smbxi}}
\def\smbomicronhat{\widehat{\smbomicron}}
\def\smbpihat{\widehat{\smbpi}}
\def\smbrhohat{\widehat{\smbrho}}
\def\smbsigmahat{\widehat{\smbsigma}}
\def\smbtauhat{\widehat{\smbtau}}
\def\smbupsilonhat{\widehat{\smbupsilon}}
\def\smbphihat{\widehat{\smbphi}}
\def\smbchihat{\widehat{\smbchi}}
\def\smbpsihat{\widehat{\smbpsi}}
\def\smbomegahat{\widehat{\smbomega}}
\def\smbAlphahat{\widehat{\smbAlpha}}
\def\smbBetahat{\widehat{\smbBeta}}
\def\smbGammahat{\widehat{\smbGamma}}
\def\smbDeltahat{\widehat{\smbDelta}}
\def\smbEpsilonhat{\widehat{\smbEpsilon}}
\def\smbZetahat{\widehat{\smbZeta}}
\def\smbEtahat{\widehat{\smbEta}}
\def\smbThetahat{\widehat{\smbTheta}}
\def\smbIotahat{\widehat{\smbIota}}
\def\smbKappahat{\widehat{\smbKappa}}
\def\smbLambdahat{\widehat{\smbLambda}}
\def\smbMuhat{\widehat{\smbMu}}
\def\smbNuhat{\widehat{\smbNu}}
\def\smbXihat{\widehat{\smbXi}}
\def\smbOmicronhat{\widehat{\smbOmicron}}
\def\smbPihat{\widehat{\smbPi}}
\def\smbRhohat{\widehat{\smbRho}}
\def\smbSigmahat{\widehat{\smbSigma}}
\def\smbTauhat{\widehat{\smbTau}}
\def\smbUpsilonhat{\widehat{\smbUpsilon}}
\def\smbPhihat{\widehat{\smbPhi}}
\def\smbChihat{\widehat{\smbChi}}
\def\smbPsihat{\widehat{\smbPsi}}
\def\smbOmegahat{\widehat{\smbOmega}}
%
%
%
%
%
\def\balphatilde{{\widetilde\balpha}}
\def\bbetatilde{{\widetilde\bbeta}}
\def\bgammatilde{{\widetilde\bgamma}}
\def\bdeltatilde{{\widetilde\bdelta}}
\def\bepsilontilde{{\widetilde\bepsilon}}
\def\bzetatilde{{\widetilde\bzeta}}
\def\bdetatilde{{\widetilde\bdeta}}
\def\bthetatilde{{\widetilde\btheta}}
\def\biotatilde{{\widetilde\biota}}
\def\bkappatilde{{\widetilde\bkappa}}
\def\blambdatilde{{\widetilde\blambda}}
\def\bmutilde{{\widetilde\bmu}}
\def\bnutilde{{\widetilde\bnu}}
\def\bxitilde{{\widetilde\bxi}}
\def\bomicrontilde{{\widetilde\bomicron}}
\def\bpitilde{{\widetilde\bpi}}
\def\brhotilde{{\widetilde\brho}}
\def\bsigmatilde{{\widetilde\bsigma}}
\def\btautilde{{\widetilde\btau}}
\def\bupsilontilde{{\widetilde\bupsilon}}
\def\bphitilde{{\widetilde\bphi}}
\def\bchitilde{{\widetilde\bchi}}
\def\bpsitilde{{\widetilde\bpsi}}
\def\bomegatilde{{\widetilde\bomega}}
\def\bAlphatilde{{\widetilde\bAlpha}}
\def\bBetatilde{{\widetilde\bBeta}}
\def\bGammatilde{{\widetilde\bGamma}}
\def\bDeltatilde{{\widetilde\bDelta}}
\def\bEpsilontilde{{\widetilde\bEpsilon}}
\def\bZetatilde{{\widetilde\bZeta}}
\def\bEtatilde{{\widetilde\bEta}}
\def\bThetatilde{{\widetilde\bTheta}}
\def\bIotatilde{{\widetilde\bIota}}
\def\bKappatilde{{\widetilde\bKappa}}
\def\bLambdatilde{{\widetilde\bLambda}}
\def\bMutilde{{\widetilde\bMu}}
\def\bNutilde{{\widetilde\bNu}}
\def\bXitilde{{\widetilde\bXi}}
\def\bOmicrontilde{{\widetilde\bOmicron}}
\def\bPitilde{{\widetilde\bPi}}
\def\bRhotilde{{\widetilde\bRho}}
\def\bSigmatilde{{\widetilde\bSigma}}
\def\bTautilde{{\widetilde\bTau}}
\def\bUpsilontilde{{\widetilde\bUpsilon}}
\def\bPhitilde{{\widetilde\bPhi}}
\def\bChitilde{{\widetilde\bChi}}
\def\bPsitilde{{\widetilde\bPsi}}
\def\bOmegatilde{{\widetilde\bOmega}}
%
%
%
%
%
\def\abar{\bar{ a}}
\def\bbar{\bar{ b}}
\def\cbar{\bar{ c}}
\def\dbar{\bar{ d}}
\def\ebar{\bar{ e}}
\def\fbar{\bar{ f}}
\def\gbar{\bar{ g}}
\def\hbar{\bar{ h}}
\def\ibar{\bar{ i}}
\def\jbar{\bar{ j}}
\def\kbar{\bar{ k}}
\def\lbar{\bar{ l}}
\def\mbar{\bar{ m}}
\def\nbar{\bar{ n}}
\def\obar{\bar{ o}}
\def\pbar{\bar{ p}}
\def\qbar{\bar{ q}}
\def\rbar{\bar{ r}}
\def\sbar{\bar{ s}}
\def\tbar{\bar{ t}}
\def\ubar{\bar{ u}}
\def\vbar{\bar{ v}}
\def\wbar{\bar{ w}}
\def\xbar{\bar{ x}}
\def\ybar{\bar{ y}}
\def\zbar{\bar{ z}}
\def\Abar{\bar{ A}}
\def\Bbar{\bar{ B}}
\def\Cbar{\bar{ C}}
\def\Dbar{\bar{ D}}
\def\Ebar{\bar{ E}}
\def\Fbar{\bar{ F}}
\def\Gbar{\bar{ G}}
\def\Hbar{\bar{ H}}
\def\Ibar{\bar{ I}}
\def\Jbar{\bar{ J}}
\def\Kbar{\bar{ K}}
\def\Lbar{\bar{ L}}
\def\Mbar{\bar{ M}}
\def\Nbar{\bar{ N}}
\def\Obar{\bar{ O}}
\def\Pbar{\bar{ P}}
\def\Qbar{\bar{ Q}}
\def\Rbar{\bar{ R}}
\def\Sbar{\bar{ S}}
\def\Tbar{\bar{ T}}
\def\Ubar{\bar{ U}}
\def\Vbar{\bar{ V}}
\def\Wbar{\bar{ W}}
\def\Xbar{\bar{ X}}
\def\Ybar{\bar{ Y}}
\def\Zbar{\bar{ Z}}
%
%
%
%
%
\def\babar{\bar{ \ba}}
\def\bbbar{\bar{ \bb}}
\def\bcbar{\bar{ \bc}}
\def\bdbar{\bar{ \bd}}
\def\bebar{\bar{ \be}}
\def\bfbar{\bar{ \bf}}
\def\bgbar{\bar{ \bg}}
\def\bhbar{\bar{ \bh}}
\def\bibar{\bar{ \bi}}
\def\bjbar{\bar{ \bj}}
\def\bkbar{\bar{ \bk}}
\def\blbar{\bar{ \bl}}
\def\bmbar{\bar{ \bm}}
\def\bnbar{\bar{ \bn}}
\def\bobar{\bar{ \bo}}
\def\bpbar{\bar{ \bp}}
\def\bqbar{\bar{ \bq}}
\def\brbar{\bar{ \br}}
\def\bsbar{\bar{ \bs}}
\def\btbar{\bar{ \bt}}
\def\bubar{\bar{ \bu}}
\def\bvbar{\bar{ \bv}}
\def\bwbar{\bar{ \bw}}
\def\bxbar{\bar{ \bx}}
\def\bybar{\bar{ \by}}
\def\bzbar{\bar{ \bz}}
\def\bAbar{\bar{ \bA}}
\def\bBbar{\bar{ \bB}}
\def\bCbar{\bar{ \bC}}
\def\bDbar{\bar{ \bD}}
\def\bEbar{\bar{ \bE}}
\def\bFbar{\bar{ \bF}}
\def\bGbar{\bar{ \bG}}
\def\bHbar{\bar{ \bH}}
\def\bIbar{\bar{ \bI}}
\def\bJbar{\bar{ \bJ}}
\def\bKbar{\bar{ \bK}}
\def\bLbar{\bar{ \bL}}
\def\bMbar{\bar{ \bM}}
\def\bNbar{\bar{ \bN}}
\def\bObar{\bar{ \bO}}
\def\bPbar{\bar{ \bP}}
\def\bQbar{\bar{ \bQ}}
\def\bRbar{\bar{ \bR}}
\def\bSbar{\bar{ \bS}}
\def\bTbar{\bar{ \bT}}
\def\bUbar{\bar{ \bU}}
\def\bVbar{\bar{ \bV}}
\def\bWbar{\bar{ \bW}}
\def\bXbar{\bar{ \bX}}
\def\bYbar{\bar{ \bY}}
\def\bZbar{\bar{ \bZ}}
%
%

%
%
%
\def\asc{{\cal a}}
\def\bsc{{\cal b}}
\def\csc{{\cal c}}
\def\dsc{{\cal d}}
\def\esc{{\cal e}}
\def\dsc{{\cal f}}
\def\gsc{{\cal g}}
\def\hsc{{\cal h}}
\def\isc{{\cal i}}
\def\jsc{{\cal j}}
\def\ksc{{\cal k}}
\def\lsc{{\cal l}}
\def\msc{{\cal m}}
\def\nsc{{\cal n}}
\def\osc{{\cal o}}
\def\psc{{\cal p}}
\def\qsc{{\cal q}}
\def\rsc{{\cal r}}
\def\ssc{{\cal s}}
\def\tsc{{\cal t}}
\def\usc{{\cal u}}
\def\vsc{{\cal v}}
\def\wsc{{\cal w}}
\def\xsc{{\cal x}}
\def\ysc{{\cal y}}
\def\zsc{{\cal z}}
\def\Asc{{\cal A}}
\def\Bsc{{\cal B}}
\def\Csc{{\cal C}}
\def\Dsc{{\cal D}}
\def\Esc{{\cal E}}
\def\Fsc{{\cal F}}
\def\Gsc{{\cal G}}
\def\Hsc{{\cal H}}
\def\Isc{{\cal I}}
\def\Jsc{{\cal J}}
\def\Ksc{{\cal K}}
\def\Lsc{{\cal L}}
\def\Msc{{\cal M}}
\def\Nsc{{\cal N}}
\def\Osc{{\cal O}}
\def\Psc{{\cal P}}
\def\Qsc{{\cal Q}}
\def\Rsc{{\cal R}}
\def\Ssc{{\cal S}}
\def\Tsc{{\cal T}}
\def\Usc{{\cal U}}
\def\Vsc{{\cal V}}
\def\Wsc{{\cal W}}
\def\Xsc{{\cal X}}
\def\Ysc{{\cal Y}}
\def\Zsc{{\cal Z}}
\def\Aschat{\widehat{{\cal A}}}
\def\Bschat{\widehat{{\cal B}}}
\def\Cschat{\widehat{{\cal C}}}
\def\Dschat{\widehat{{\cal D}}}
\def\Eschat{\widehat{{\cal E}}}
\def\Fschat{\widehat{{\cal F}}}
\def\Gschat{\widehat{{\cal G}}}
\def\Hschat{\widehat{{\cal H}}}
\def\Ischat{\widehat{{\cal I}}}
\def\Jschat{\widehat{{\cal J}}}
\def\Kschat{\widehat{{\cal K}}}
\def\Lschat{\widehat{{\cal L}}}
\def\Mschat{\widehat{{\cal M}}}
\def\Nschat{\widehat{{\cal N}}}
\def\Oschat{\widehat{{\cal O}}}
\def\Pschat{\widehat{{\cal P}}}
\def\Qschat{\widehat{{\cal Q}}}
\def\Rschat{\widehat{{\cal R}}}
\def\Sschat{\widehat{{\cal S}}}
\def\Tschat{\widehat{{\cal T}}}
\def\Uschat{\widehat{{\cal U}}}
\def\Vschat{\widehat{{\cal V}}}
\def\Wschat{\widehat{{\cal W}}}
\def\Xschat{\widehat{{\cal X}}}
\def\Yschat{\widehat{{\cal Y}}}
\def\Zschat{\widehat{{\cal Z}}}
\def\Asctilde{\widetilde{{\cal A}}}
\def\Bsctilde{\widetilde{{\cal B}}}
\def\Csctilde{\widetilde{{\cal C}}}
\def\Dsctilde{\widetilde{{\cal D}}}
\def\Esctilde{\widetilde{{\cal E}}}
\def\Fsctilde{\widetilde{{\cal F}}}
\def\Gsctilde{\widetilde{{\cal G}}}
\def\Hsctilde{\widetilde{{\cal H}}}
\def\Isctilde{\widetilde{{\cal I}}}
\def\Jsctilde{\widetilde{{\cal J}}}
\def\Ksctilde{\widetilde{{\cal K}}}
\def\Lsctilde{\widetilde{{\cal L}}}
\def\Msctilde{\widetilde{{\cal M}}}
\def\Nsctilde{\widetilde{{\cal N}}}
\def\Osctilde{\widetilde{{\cal O}}}
\def\Psctilde{\widetilde{{\cal P}}}
\def\Qsctilde{\widetilde{{\cal Q}}}
\def\Rsctilde{\widetilde{{\cal R}}}
\def\Ssctilde{\widetilde{{\cal S}}}
\def\Tsctilde{\widetilde{{\cal T}}}
\def\Usctilde{\widetilde{{\cal U}}}
\def\Vsctilde{\widetilde{{\cal V}}}
\def\Wsctilde{\widetilde{{\cal W}}}
\def\Xsctilde{\widetilde{{\cal X}}}
\def\Ysctilde{\widetilde{{\cal Y}}}
\def\Zsctilde{\widetilde{{\cal Z}}}
\def\bAsc{\mathbf{\cal A}}
\def\bBsc{\mathbf{\cal B}}
\def\bCsc{\mathbf{\cal C}}
\def\bDsc{\mathbf{\cal D}}
\def\bEsc{\mathbf{\cal E}}
\def\bFsc{\mathbf{\cal F}}
\def\bGsc{\mathbf{\cal G}}
\def\bHsc{\mathbf{\cal H}}
\def\bIsc{\mathbf{\cal I}}
\def\bJsc{\mathbf{\cal J}}
\def\bKsc{\mathbf{\cal K}}
\def\bLsc{\mathbf{\cal L}}
\def\bMsc{\mathbf{\cal M}}
\def\bNsc{\mathbf{\cal N}}
\def\bOsc{\mathbf{\cal O}}
\def\bPsc{\mathbf{\cal P}}
\def\bQsc{\mathbf{\cal Q}}
\def\bRsc{\mathbf{\cal R}}
\def\bSsc{\mathbf{\cal S}}
\def\bTsc{\mathbf{\cal T}}
\def\bUsc{\mathbf{\cal U}}
\def\bVsc{\mathbf{\cal V}}
\def\bWsc{\mathbf{\cal W}}
\def\bXsc{\mathbf{\cal X}}
\def\bYsc{\mathbf{\cal Y}}
\def\bZsc{\mathbf{\cal Z}}
\def\bAschat{\widehat{\mathbf{\cal A}}}
\def\bBschat{\widehat{\mathbf{\cal B}}}
\def\bCschat{\widehat{\mathbf{\cal C}}}
\def\bDschat{\widehat{\mathbf{\cal D}}}
\def\bEschat{\widehat{\mathbf{\cal E}}}
\def\bFschat{\widehat{\mathbf{\cal F}}}
\def\bGschat{\widehat{\mathbf{\cal G}}}
\def\bHschat{\widehat{\mathbf{\cal H}}}
\def\bIschat{\widehat{\mathbf{\cal I}}}
\def\bJschat{\widehat{\mathbf{\cal J}}}
\def\bKschat{\widehat{\mathbf{\cal K}}}
\def\bLschat{\widehat{\mathbf{\cal L}}}
\def\bMschat{\widehat{\mathbf{\cal M}}}
\def\bNschat{\widehat{\mathbf{\cal N}}}
\def\bOschat{\widehat{\mathbf{\cal O}}}
\def\bPschat{\widehat{\mathbf{\cal P}}}
\def\bQschat{\widehat{\mathbf{\cal Q}}}
\def\bRschat{\widehat{\mathbf{\cal R}}}
\def\bSschat{\widehat{\mathbf{\cal S}}}
\def\bTschat{\widehat{\mathbf{\cal T}}}
\def\bUschat{\widehat{\mathbf{\cal U}}}
\def\bVschat{\widehat{\mathbf{\cal V}}}
\def\bWschat{\widehat{\mathbf{\cal W}}}
\def\bXschat{\widehat{\mathbf{\cal X}}}
\def\bYschat{\widehat{\mathbf{\cal Y}}}
\def\bZschat{\widehat{\mathbf{\cal Z}}}
\def\afrak{\mathfrak{a}}
\def\bfrak{\mathfrak{b}}
\def\cfrak{\mathfrak{c}}
\def\dfrak{\mathfrak{d}}
\def\efrak{\mathfrak{e}}
\def\ffrak{\mathfrak{f}}
\def\gfrak{\mathfrak{g}}
\def\hfrak{\mathfrak{h}}
\def\ifrak{\mathfrak{i}}
\def\jfrak{\mathfrak{j}}
\def\kfrak{\mathfrak{k}}
\def\lfrak{\mathfrak{l}}
\def\mfrak{\mathfrak{m}}
\def\nfrak{\mathfrak{n}}
\def\ofrak{\mathfrak{o}}
\def\pfrak{\mathfrak{p}}
\def\qfrak{\mathfrak{q}}
\def\rfrak{\mathfrak{r}}
\def\sfrak{\mathfrak{s}}
\def\tfrak{\mathfrak{t}}
\def\ufrak{\mathfrak{u}}
\def\vfrak{\mathfrak{v}}
\def\wfrak{\mathfrak{w}}
\def\xfrak{\mathfrak{x}}
\def\yfrak{\mathfrak{y}}
\def\zfrak{\mathfrak{z}}
\def\Afrak{\mathfrak{ A}}
\def\Bfrak{\mathfrak{ B}}
\def\Cfrak{\mathfrak{ C}}
\def\Dfrak{\mathfrak{ D}}
\def\Efrak{\mathfrak{ E}}
\def\Ffrak{\mathfrak{ F}}
\def\Gfrak{\mathfrak{ G}}
\def\Hfrak{\mathfrak{ H}}
\def\Ifrak{\mathfrak{ I}}
\def\Jfrak{\mathfrak{ J}}
\def\Kfrak{\mathfrak{ K}}
\def\Lfrak{\mathfrak{ L}}
\def\Mfrak{\mathfrak{ M}}
\def\Nfrak{\mathfrak{ N}}
\def\Ofrak{\mathfrak{ O}}
\def\Pfrak{\mathfrak{ P}}
\def\Qfrak{\mathfrak{ Q}}
\def\Rfrak{\mathfrak{ R}}
\def\Sfrak{\mathfrak{ S}}
\def\Tfrak{\mathfrak{ T}}
\def\Ufrak{\mathfrak{ U}}
\def\Vfrak{\mathfrak{ V}}
\def\Wfrak{\mathfrak{ W}}
\def\Xfrak{\mathfrak{ X}}
\def\Yfrak{\mathfrak{ Y}}
\def\Zfrak{\mathfrak{ Z}}

\def\bAfrak{\mathbf{\mathfrak{A}}}
\def\bBfrak{\mathbf{\mathfrak{B}}}
\def\bCfrak{\mathbf{\mathfrak{C}}}
\def\bDfrak{\mathbf{\mathfrak{D}}}
\def\bEfrak{\mathbf{\mathfrak{E}}}
\def\bFfrak{\mathbf{\mathfrak{F}}}
\def\bGfrak{\mathbf{\mathfrak{G}}}
\def\bHfrak{\mathbf{\mathfrak{H}}}
\def\bIfrak{\mathbf{\mathfrak{I}}}
\def\bJfrak{\mathbf{\mathfrak{J}}}
\def\bKfrak{\mathbf{\mathfrak{K}}}
\def\bLfrak{\mathbf{\mathfrak{L}}}
\def\bMfrak{\mathbf{\mathfrak{M}}}
\def\bNfrak{\mathbf{\mathfrak{N}}}
\def\bOfrak{\mathbf{\mathfrak{O}}}
\def\bPfrak{\mathbf{\mathfrak{P}}}
\def\bQfrak{\mathbf{\mathfrak{Q}}}
\def\bRfrak{\mathbf{\mathfrak{R}}}
\def\bSfrak{\mathbf{\mathfrak{S}}}
\def\bTfrak{\mathbf{\mathfrak{T}}}
\def\bUfrak{\mathbf{\mathfrak{U}}}
\def\bVfrak{\mathbf{\mathfrak{V}}}
\def\bWfrak{\mathbf{\mathfrak{W}}}
\def\bXfrak{\mathbf{\mathfrak{X}}}
\def\bYfrak{\mathbf{\mathfrak{Y}}}
\def\bZfrak{\mathbf{\mathfrak{Z}}}

\def\bAfrakhat{\mathbf{\widehat{\mathfrak{A}}}}
\def\bBfrakhat{\mathbf{\widehat{\mathfrak{B}}}}
\def\bCfrakhat{\mathbf{\widehat{\mathfrak{C}}}}
\def\bDfrakhat{\mathbf{\widehat{\mathfrak{D}}}}
\def\bEfrakhat{\mathbf{\widehat{\mathfrak{E}}}}
\def\bFfrakhat{\mathbf{\widehat{\mathfrak{F}}}}
\def\bGfrakhat{\mathbf{\widehat{\mathfrak{G}}}}
\def\bHfrakhat{\mathbf{\widehat{\mathfrak{H}}}}
\def\bIfrakhat{\mathbf{\widehat{\mathfrak{I}}}}
\def\bJfrakhat{\mathbf{\widehat{\mathfrak{J}}}}
\def\bKfrakhat{\mathbf{\widehat{\mathfrak{K}}}}
\def\bLfrakhat{\mathbf{\widehat{\mathfrak{L}}}}
\def\bMfrakhat{\mathbf{\widehat{\mathfrak{M}}}}
\def\bNfrakhat{\mathbf{\widehat{\mathfrak{N}}}}
\def\bOfrakhat{\mathbf{\widehat{\mathfrak{O}}}}
\def\bPfrakhat{\mathbf{\widehat{\mathfrak{P}}}}
\def\bQfrakhat{\mathbf{\widehat{\mathfrak{Q}}}}
\def\bRfrakhat{\mathbf{\widehat{\mathfrak{R}}}}
\def\bSfrakhat{\mathbf{\widehat{\mathfrak{S}}}}
\def\bTfrakhat{\mathbf{\widehat{\mathfrak{T}}}}
\def\bUfrakhat{\mathbf{\widehat{\mathfrak{U}}}}
\def\bVfrakhat{\mathbf{\widehat{\mathfrak{V}}}}
\def\bWfrakhat{\mathbf{\widehat{\mathfrak{W}}}}
\def\bXfrakhat{\mathbf{\widehat{\mathfrak{X}}}}
\def\bYfrakhat{\mathbf{\widehat{\mathfrak{Y}}}}
\def\bZfrakhat{\mathbf{\widehat{\mathfrak{Z}}}}
%
%
%
%
\def\etal{{\em et al.}}
%
%
%
%
%
\def\cumsum{\mbox{cumsum}}
\def\real{{\mathbb R}}
\def\intinfinf{\int_{-\infty}^{\infty}}
\def\intzinf{\int_{0}^{\infty}}
\def\intzt{\int_0^t}
\def\transpose{{\sf \scriptscriptstyle{T}}}
\def\smhalf{{\textstyle{1\over2}}}
\def\third{{\textstyle{1\over3}}}
\def\twothirds{{\textstyle{2\over3}}}
\def\bell{\bmath{\ell}}
\def\half{\frac{1}{2}}
\def\ninv{n^{-1}}
\def\nhalf{n^{\half}}
\def\mhalf{m^{\half}}
\def\nnhalf{n^{-\half}}
\def\mnhalf{m^{-\half}}
\def\MN{\mbox{MN}}
\def\N{\mbox{N}}
\def\E{\mbox{E}}
\def\pr{P}
\def\var{\mbox{var}}
\def\limn{\lim_{n\to \infty} }
\def\intt{\int_{\tau_a}^{\tau_b}}
\def\sumin{\sum_{i=1}^n}
\def\sumjn{\sum_{j=1}^n}
\def\SUMin{{\displaystyle \sum_{i=1}^n}}
\def\SUMjn{{\displaystyle \sum_{j=1}^n}}
\def\myendthm{\begin{flushright} $\diamond $ \end{flushright}}
\def\convd{\overset{\Dsc}{\longrightarrow}}
\def\convp{\overset{\Psc}{\longrightarrow}}
\def\convas{\overset{a.s.}{\longrightarrow}}
\def\hn{\mbox{H}_0}
\def\ha{\mbox{H}_1}

%
%
%
%
%
\def\trans{^{\transpose}}
\def\inv{^{-1}}
\def\twobyone#1#2{\left[
\begin{array}
{c}
#1\\
#2\\
\end{array}
\right]}
%
%
%
%
%
\def\argmindum{\mathop{\mbox{argmin}}}
\def\argmin#1{\argmindum_{#1}}
\def\argmaxdum{\mathop{\mbox{argmax}}}
\def\argmax#1{\argmaxdum_{#1}}
\def\blockdiag{\mbox{blockdiag}}
\def\diag{\mbox{diag}}
\def\dffit{df_{{\rm fit}}}
\def\dfres{df_{{\rm res}}}
\def\dfyhat{df_{\yhat}}
\def\diag{\mbox{diag}}
\def\diagonal{\mbox{diagonal}}
\def\logit{\mbox{logit}}
\def\stdev{\mbox{st.\,dev.}}
\def\stdevhat{{\widehat{\mbox{st.dev}}}}
\def\tr{\mbox{tr}}
\def\trigamma{\mbox{trigamma}}
\def\vecof{\mbox{vec}}
\def\AIC{\mbox{AIC}}
\def\AMISE{\mbox{AMISE}}
\def\corr{\mbox{corr}}
\def\cov{\mbox{cov}}
\def\Corr{\mbox{Corr}}
\def\Cov{\mbox{Cov}}
\def\Var{\mbox{Var}}
\def\var{\mbox{var}}
\def\CV{\mbox{CV}}
\def\GCV{\mbox{GCV}}
\def\LR{\mbox{LR}}
\def\MISE{\mbox{MISE}}
\def\MSSE{\mbox{MSSE}}
\def\ML{\mbox{ML}}
\def\REML{\mbox{REML}}
\def\RMSE{{\rm RMSE}}
\def\RSS{\mbox{RSS}}
%
%
%
%
\def\bib{\vskip12pt\par\noindent\hangindent=1 true cm\hangafter=1}
\def\jump{\vskip3mm\noindent}
\def\mybox#1{\vskip1mm \begin{center}
        \hspace{.0\textwidth}\vbox{\hrule\hbox{\vrule\kern6pt
\parbox{.9\textwidth}{\kern6pt#1\vskip6pt}\kern6pt\vrule}\hrule}
        \end{center} \vskip-5mm}
\def\lboxit#1{\vbox{\hrule\hbox{\vrule\kern6pt
      \vbox{\kern6pt#1\vskip6pt}\kern6pt\vrule}\hrule}}
\def\boxit#1{\begin{center}\fbox{#1}\end{center}}
\def\thickboxit#1{\vbox{{\hrule height 1mm}\hbox{{\vrule width 1mm}\kern6pt
          \vbox{\kern6pt#1\kern6pt}\kern6pt{\vrule width 1mm}}
               {\hrule height 1mm}}}
\def\instep{\vskip12pt\par\hangindent=30 true mm\hangafter=1}
\def\uWand{\underline{Wand}}
\def\remtask#1#2{\mmnote{\thickboxit
                 {\bf #1\ \theremtask}}\refstepcounter{remtask}}
%
%
%

%
%
\def\aism{{\it Ann. Inst. Statist. Math.}\ }
\def\ajs{{\it Austral. J. Statist.}\ }
\def\ANNSTAT{{\it The Annals of Statistics}\ }
\def\annmath{{\it Ann. Math. Statist.}\ }
\def\applstat{{\it Appl. Statist.}\ }
\def\BIOMETRICS{{\it Biometrics}\ }
\def\cjs{{\it Canad. J. Statist.}\ }
\def\csda{{\it Comp. Statist. Data Anal.}\ }
\def\cstm{{\it Comm. Statist. Theory Meth.}\ }
\def\ieeetit{{\it IEEE Trans. Inf. Theory}\ }
\def\isr{{\it Internat. Statist. Rev.}\ }
\def\JASA{{\it Journal of the American Statistical Association}\ }
\def\JCGS{{\it Journal of Computational and Graphical Statistics}\ }
\def\jscs{{\it J. Statist. Comput. Simulation}\ }
\def\jma{{\it J. Multivariate Anal.}\ }
\def\jns{{\it J. Nonparametric Statist.}\ }
\def\JRSSA{{\it Journal of the Royal Statistics Society, Series A}\ }
\def\JRSSB{{\it Journal of the Royal Statistics Society, Series B}\ }
\def\JRSSC{{\it Journal of the Royal Statistics Society, Series C}\ }
\def\jspi{{\it J. Statist. Planning Inference}\ }
\def\ptrf{{\it Probab. Theory Rel. Fields}\ }
\def\sankhyaa{{\it Sankhy$\bar{{\it a}}$} Ser. A\ }
\def\sjs{{\it Scand. J. Statist.}\ }
\def\spl{{\it Statist. Probab. Lett.}\ }
\def\statsci{{\it Statist. Sci.}\ }
\def\techno{{\it Technometrics}\ }
\def\tpa{{\it Theory Probab. Appl.}\ }
\def\zw{{\it Z. Wahr. ver. Geb.}\ }
%
%
%
%
\def\Brent{{\bf BRENT:}\ }
\def\David{{\bf DAVID:}\ }
\def\Erin{{\bf ERIN:}}
\def\Gerda{{\bf GERDA:}\ }
\def\Joel{{\bf JOEL:}\ }
\def\Marc{{\bf MARC:}\ }
\def\Matt{{\bf MATT:}\ }
\def\Tianxi{{\bf TIANXI:}\ }
%
%
%
%
\def\bZE{\bZ_{\scriptscriptstyle E}}
\def\bZT{\bZ_{\scriptscriptstyle T}}
\def\bbE{\bb_{\scriptscriptstyle E}}
\def\bbT{\bb_{\scriptscriptstyle T}}
\def\bbhatT{\bbhat_{\scriptscriptstyle T}}
\def\fX{f_{\scriptscriptstyle X}}
\def\sigeps{\sigma_{\varepsilon}}
\def\bVtheta{\bV_{\smbtheta}}
\def\bVthetainv{\bVtheta^{-1}}
\def\bKsc{\boldsymbol{\Ksc}}
\def\bxbar{\bar{\bx}}
\def\bPL{b^{\scriptscriptstyle{\rm PL}}}
\def\bVA{b^{\scriptscriptstyle{\rm VA}}}
\def\zPL{z^{\scriptscriptstyle{\rm PL}}}
\def\zVA{z^{\scriptscriptstyle{\rm VA}}}
\def\bYmis{\bY_{\scriptscriptstyle{\rm mis}}}
\def\bYmishat{{\widehat{\bYmis}}}
\def\bYmisone{\bY_{\scriptscriptstyle{\rm mis,1}}}
\def\bYmistwo{\bY_{\scriptscriptstyle{\rm mis,2}}}
\def\bYobs{\bY_{\scriptscriptstyle{\rm obs}}}
\def\bdobs{\bd_{\scriptscriptstyle{\rm obs}}}
\def\bdmis{\bd_{\scriptscriptstyle{\rm mis}}}
%
%
%
%
\def\bfDelta{{\mbox{\boldmath$\Delta$}}}
\def\bfkappa{{\mbox{\boldmath$\kappa$}}}
\def\bfgamma{{\mbox{\boldmath$\gamma$}}}
\def\bftheta{{\mbox{\boldmath$\theta$}}}
\def\bfmu{{\mbox{\boldmath$\mu$}}}
\def\bfdelta{{\mbox{\boldmath$\delta$}}}
\def\bfeps{{\mbox{\boldmath$\varepsilon$}}}
\def\bfnu{{\mbox{\boldmath$\nu$}}}
\def\bfzeta{{\mbox{\boldmath$\zeta$}}}
\def\bfchi{{\mbox{\boldmath$\chi$}}}
\def\bbX{\mathbb{X}}
\def\bbV{\mathbb{V}} 
\def\bbA{\mathbb{A}}
\def\bbB{\mathbb{B}}
\def\bbE{\mathbb{E}}
\def\mpr{\mathbb{P}}
\def\mfdr{\mbox{mFDR}}
\def\fdr{\mbox{FDR}}
\def\tdr{\mbox{TDR}}

%
%
%
%
\def\miss{\mbox{{\tiny miss}}}
\def\obs{\mbox{{\tiny obs}}}

%
%
%
%
\def\bmath#1{\mbox{\boldmath$#1$}}
\def\fat#1{\hbox{\rlap{$#1$}\kern0.25pt\rlap{$#1$}\kern0.25pt$#1$}}
\def\wh{\widehat}
\def\flambda{\fat{\lambda}}
\def\beps{\bmath{\varepsilon}}
\def\bSlambda{\bS_{\lambda}}
\def\ErrorSS{\mbox{RSS}}
\def\bsqbar{\bar{{b^2}}}
\def\bcubar{\bar{{b^3}}}
\def\plargest{p_{\rm \,largest}}
\def\summheading#1{\subsection*{#1}\hskip3mm}
\def\summbreak{\vskip3mm\par}
\def\df{df}
\def\adf{adf}
\def\dffit{df_{{\rm fit}}}
\def\dfres{df_{{\rm res}}}
\def\dfyhat{df_{\yhat}}
\def\sigb{\sigma_b}
\def\sigu{\sigma_u}
\def\sigepshat{{\widehat\sigma}_{\varepsilon}}
\def\siguhat{{\widehat\sigma}_u}
\def\sigepshat{{\widehat\sigma}_{\varepsilon}}
\def\sigbhat{{\widehat\sigma}_b}
\def\sighat{{\widehat\sigma}}
\def\sigsqb{\sigma^2_b}
\def\sigsqeps{\sigma^2_{\varepsilon}}
\def\sigsqepszerohat{{\widehat\sigma}^2_{\varepsilon,0}}
\def\sigsqepshat{{\widehat\sigma}^2_{\varepsilon}}
\def\sigsqbhat{{\widehat\sigma}^2_b}
\def\dfnumer{{\rm df(II}|{\rm I)}}
\def\mhatlam{{\widehat m}_{\lambda}}
\def\calD{\Dsc}
\def\Aeps{A_{\epsilon}}
\def\Beps{B_{\epsilon}}
\def\Ab{A_b}
\def\Bb{B_b}
\def\bXtmain{\tilde{\bX}_r}
\def\main{\mbox{\tt main}}
\def\argminbetab{\argmin{\bbeta,\bb}}
\def\calB{\Bsc}
\def\respvar{\mbox{\tt log(amt)}}

\def\Abb{\mathbb{A}}
\def\Bbbb{\mathbb{B}}
\def\Zbb{\mathbb{Z}}
\def\Wbb{\mathbb{W}}

\def\Abbhat{\widehat{\mathbb{A}}}

\def\pn{\phantom{-}}
\def\pp{\phantom{1}}

\def\diag{\mbox{\em diag}}
\def\sumin{\sum_{i=1}^{n}}
\def\sumjp{\sum_{j=1}^{p}}
\def\convprob{\overset{\Psc}{\longrightarrow}}
\def\convdist{\overset{\Dsc}{\longrightarrow}}
\def\lasso{\mbox{\tiny LASSO}}
\def\alasso{\mbox{\tiny ALASSO}}
\def\nng{\mbox{\tiny NNG}}
\def\ridge{\mbox{\tiny ridge}}
\def\ols{\mbox{\tiny OLS}}

\def\trace{\mbox{trace}}
\def\ellhat{\widehat{\ell}}
\def\elltilde{\tilde{\ell}}
\def\bvarepsilonhat{{\widehat\bvarepsilon}}
\def\sgn{\mbox{sgn}}
\def\mle{\mbox{\tiny MLE}}

%
%

\maketitle

\begin{abstract}
    Correlation networks are commonly used to infer associations between microbes and metabolites. The resulting $p$-values are then corrected for multiple comparisons using existing methods such as the Benjamini \& Hochberg (BH) procedure to control the false discovery rate (FDR). However, most existing methods for FDR control assume the $p$-values are weakly dependent. Consequently, they can have low power in recovering microbe--metabolite association networks that exhibit important topological features, such as the presence of densely associated modules. We propose a novel inference procedure that is both powerful for detecting significant associations in the microbe--metabolite network and capable of controlling the FDR. Power enhancement is achieved by modeling latent structures in the form of a bipartite stochastic block model. We develop a variational expectation–maximization (EM) algorithm to estimate the model parameters and incorporate the learned graph in the testing procedure. In addition to FDR control, this procedure provides a clustering of microbes and metabolites into modules, which is useful for interpretation. We demonstrate the merit of the proposed method in simulations and an application to bacterial vaginosis. 
\end{abstract}


 
\newpage
\doublespacing

\section{Introduction}

The gut microbiome plays an important role in human health and diseases \cite{hou2022microbiota}. Because the gut microbiome is amenable to interventions, it has great translational potential to be used clinically for addressing various health conditions \cite{gebrayel2022microbiota}. One potential pathway by which the gut microbiome impacts host health is through microbial-derived metabolites: these molecules can modulate the immune system  \cite{dinglasan2025microbial,yaqub2025microbiome,williams2024harnessing}. Yet, despite extensive experimental and computational work, a large number of microbe--metabolite relationships remain unknown, leaving a significant knowledge gap \cite{liu2022functions}. 

\subsection{Motivating study}
Our motivating example concerns a study of bacterial vaginosis (BV) using paired microbiome and metabolome profiling \cite{mcmillan2015multi}. The vaginal microbiota is an intricate ecosystem in which microbes trade nutrients, defense molecules, and signaling compounds with both the host epithelium and one another. Disruption of this ecosystem manifests clinically as BV, a condition affecting up to one-third of women worldwide \cite{koumans2007prevalence} and linked to adverse outcomes ranging from pre-term birth to heightened susceptibility to sexually transmitted infections \cite{guerra2006pregnancy,atashili2008bacterial}. A growing body of evidence suggests that BV is not driven by a single pathogen but reflect a community-wide metabolic shift \cite{srinivasan2015metabolic,laniewski2021bacterial}. Improved diagnosis and treatment of BV requires a detailed understanding of these metabolic activities. 

The data set contains 49 taxa and 128 metabolites collected from 131 individuals (see details on preprocessing in Section \ref{sec:app}). For each pair of taxon and metabolite, a test statistic, e.g., Spearman correlation coefficient, and its \(p\)-value can be derived. The task is to perform simultaneous inference for the null hypothesis of no association between the $i$-th taxon and the $j$-th metabolite, for \(i=1,\ldots,49\) and \(j=1,\ldots,128\). 

\subsection{The need for matrix-aware model}

The goal of large-scale hypothesis testing is to identify as many interesting ones as possible while keeping the error rate low. The false discovery rate (FDR)---defined as the average proportion of errors among the discoveries---is a commonly used metric to assess the significance of testing multiple hypotheses simultaneously. In practice, the Benjamini \& Hochberg (BH) procedure based on ranking \(p\)-values \cite{benjamini1995controlling} is often used to control FDR. While easy to implement, this vector-based approach suffers from two main limitations. First, it provides a list of significant associations that can be hard to interpret biologically. Second, BH and related methods such as Storey's \(q\)-value \cite{storey2004strong} assume weakly dependent \(p\)-values; this assumption is often violated when analyzing high-dimensional microbiome and metabolomic data, especially when microbes and metabolites form densely connected modules.

For vector-based testing problems (e.g., a list of $p$-values or $z$-values from genome-wide association studies), many authors have improved power by estimating the null/alternative mixture distributions \cite{efron2001empirical,efron2004large, sun2007oracle} and modeling the dependence explicitly \cite{liu2016multiple,sun2009large,li2024large,wang2024large}. 
Those advances, however, do not translate directly to matrix-valued data such as correlation or cross-correlation matrices: here the tests exhibit structured dependence that is not exploited by existing multiple-testing tools. In large-scale testing of correlations, \cite{liu2013gaussian,cai2016large} established valid FDR control for test statistics obtained from high-dimensional data under mild regularity conditions, but they do not take into account the structural information in the data.   

Rebafka et al. \cite{rebafka2022powerful} recently bridged this gap with the \emph{noisy stochastic block model} (noisySBM), which treats the constellation of null and non-null hypotheses itself as an SBM and models the observed statistics as a perturbed version of that latent graph. Applied naively to bipartite data, however, the standard noisySBM (i) is computationally heavy and (ii) tends to merge features of different types into the same community \cite{larremore2014efficiently}, degrading FDR control. 

\subsection{Our contributions}

We introduce a bipartite noisy SBM that models the hidden microbe–metabolite interaction network as a bipartite stochastic block model (biSBM) and embeds it in the observed statistics through a two-component mixture of the null and alternative densities. The proposed framework offers three key advances. First, separate block memberships are learned for microbes and metabolites, respecting their asymmetry and avoiding the mixing issue of the standard SBM. These inferred modules are useful for biological interpretation. Second, within each block the method pools information to compute structured \(\ell\)-values which are the posterior of a null hypothesis being true given observed data and the latent graph; thresholding these quantities controls the (marginal) FDR at the nominal level while markedly boosting the true discovery rate (TDR). Lastly, we develop a variational expectation-maximization (VEM) algorithm to estimate the model parameters. This algorithm solves two smaller, type-specific sub-problems, delivering order-of-magnitude speed-ups over the original noisySBM without sacrificing accuracy. Recently, several machine learning (ML) approaches were proposed to predict microbe--metabolite interactions with neural networks or other black-box models \cite{mallick2019predictive,morton2019learning,shtossel2024gut,reiman2021mimenet}. Compared to these methods, our method provides (i) rigorous uncertainty quantification via FDR control, and (ii) explicit module-level summary statistics that are helpful for generating testable biological hypotheses. 

Through extensive simulations we show that the bipartite noisySBM achieves superior FDR control and power compared with BH \cite{benjamini1995controlling}, Storey’s \(q\)-value \cite{storey2004strong}, the Sun \& Cai procedure \cite{sun2007oracle}, and the unmodified noisySBM \cite{rebafka2022powerful}. We then apply the method to paired microbiome and metabolomic data from 131 Rwandan women \cite{mcmillan2015multi}. The analysis uncovers coherent microbe--metabolite modules associated with BV---for instance, a block comprising Leptotrichia and Sneathia that jointly drive elevations in 2-hydroxyisovalerate and succinate, and a Lactobacillus-dominated block whose metabolic signature is suppressed in BV. These findings generate concrete hypotheses about metabolic alterations that can be pursued experimentally.

The rest of the paper is organized as follow. Section \ref{sec:method} introduces the bipartite noisySBM and the multiple testing procedure. Section \ref{sec:estimation} describes the algorithm and model selection strategy. 
Section \ref{sec:experiments} benchmarks the proposed method on simulated data and Section \ref{sec:app} applies it to BV multi-omics data. We close with a discussion in Section \ref{sec:diss}. 
\section{Bipartite graph inference }\label{sec:method}

\subsection{Noisy bipartite stochastic block model}
Let $A\in\{0,1\}^{n_1\times n_2}$ denote the unobserved bipartite adjacency matrix linking $n_1$ microbes (rows) to $n_2$ metabolites (columns).  We model $A$ with a bipartite stochastic block model (biSBM) characterized by two independent membership vectors $Z_1=(Z_{i,1})_{1\le i\le n_1}$ and $Z_2=(Z_{j,2})_{1\le j\le n_2}$ where
\[
Z_{i,1}\sim\text{Multinomial}\bigl(1,\balpha_1\bigr),\qquad
Z_{j,2}\sim\text{Multinomial}\bigl(1,\balpha_2\bigr).
\]
The mixing proportions
$\balpha_1=(\alpha_{q,1})_{1\le q\le B_1}$ and
$\balpha_2=(\alpha_{l,2})_{1\le q\le B_2}$ satisfy $\sum_{q=1}^{{B_1}}\alpha_{q,1} = 1$ and $\sum_{q=1}^{{{B_2}}}\alpha_{q,2} = 1$.
Conditional on the membership vectors, the edges are independent Bernoulli variables with
\[
A_{ij}\;|\;Z_1,Z_2\;\sim\;\text{Bernoulli}\bigl(\pi_{Z_{i,1},\,Z_{j,2}}\bigr),
\qquad
\Pi=(\pi_{q\ell})\in(0,1)^{B_1\times B_2}.
\]
For each pair $(i,j)$ we observe a test statistic $x_{ij}$:
\[
x_{ij}\,\bigl|\,A,Z_1,Z_2 \sim
\begin{cases}
g_0(\,\cdot\,;\nu_0), & A_{ij}=0,\\
g(\,\cdot\,;\nu_{Z_{i,1},\,Z_{j,2}}), & A_{ij}=1,
\end{cases}
\]
where $g_0(\cdot; \nu_0)$ and $g(\cdot; \nu)$ are pre-specified null and alternative densities with parameters $\nu_0\subset \Vsc_0$ and $\nu\subset \Vsc$. Both families are parametric with \(\Vsc_0 \subset \mathbb R^{d_0}\) and \(\Vsc\subset \mathbb R^{d_1}\), where \(d_0\) and \(d_1\) are the respective dimensions of the parameter spaces. The density functions $g_0$ and $g$ are usually Gaussian but can take other forms. The complete parameter set is $\theta=(\balpha_1,\balpha_2,\Pi,\nu_0,\nu)$.

The formulation reduces to the noisySBM of Rebafka–Roquain–Villers \cite{rebafka2022powerful} when $B_1\!=\!B_2$ and rows/columns share a single clustering, but the bipartite version avoids the tendency of a standard SBM to \emph{merge} different vertex types into the same community and greatly improves both speed and accuracy for bipartite data \cite{larremore2014efficiently}. In general, the number of row clusters ${B_1}$ and the number of column clusters ${{B_2}}$ do not have to be the same \cite{yen2020community}.

\subsection{Identifiability}

Before we discuss the testing procedure and model estimation, we address the model's identifiability issue. To this end, we introduce the following assumptions. 

\begin{assumption}\label{ass:distinct}
    The parameters \(\{\nu_0\subset \Vsc_0 \subset \Vsc, \nu_{ql}\subset \Vsc: 1\le q \le B_1, 1\le l \le B_2\}\) are distinct. 
\end{assumption}

\begin{assumption}\label{ass:finite}
    The parameters of any finite mixture of distributions \(\{g(\cdot; \nu): \nu \subset \Vsc\}\) are identifiable, up to label swapping. 
\end{assumption}

Assumption \ref{ass:finite} is satisfied for the class of Gaussian densities that often arises in hypothesis testing. 

\begin{theorem}\label{thm:iden}
 Let \(n_1,n_2\ge 2\) and $B_1,B_2\ge 2$. Let \(g_0(\cdot;\nu)\) and \(g(\cdot;\nu)\) be from the same family of distributions. Under Assumptions \ref{ass:distinct}-\ref{ass:finite}, all parameters of the bipartite noisy stochastic block model are identifiable, up to label swapping. 
\end{theorem}
The proof of Theorem \ref{thm:iden} can be adapted from the proofs in \cite{allman2011parameter,rebafka2022powerful} with the key modification of replacing the triangle \((X_{ij},X_{jk},X_{ki})\) with the smallest fully connected subgraph \((X_{i_1 j_1},X_{i_1 j_2},X_{i_2 j_1}, X_{i_2 j_2})\) and is thus omitted. 
When \(g(\cdot,\nu_0)\) and \(g(\cdot,\nu)\) are Gaussian densities, the model is identifiable provided the
parameter pairs $(\mu_{q\ell},\sigma_{q\ell}^2)$ and $(0,\sigma_0^2)$ are all distinct. More generally, identifiability still holds if a single alternative density is shared across blocks ($\mu_{q\ell}=\mu,\;\sigma_{q\ell}=\sigma$). 

\subsection{Graph-based multiple-testing}
Let $\varphi(X)\in\{0,1\}^{A}$ be a decision rule that rejects $H_{0,ij}$ when $\varphi_{ij}(X)=1$.  Under parameter $\theta$ we evaluate the false discovery rate (FDR), true discovery rate (TDR), and marginal FDR (mFDR) defined by
\begin{align}
\text{FDR}_{\theta}(\varphi) &=
\mathbb{E}_{\theta}\!\left[
  \frac{\sum_{(i,j)}(1-A_{ij})\varphi_{ij}(X)}
       {\bigl(\sum_{(i,j)}\varphi_{ij}(X)\bigr)\vee1}
\right],\\[4pt]
\text{TDR}_{\theta}(\varphi) &=
\frac{\mathbb{E}_{\theta}\bigl[\sum_{(i,j)}A_{ij}\varphi_{ij}(X)\bigr]}
     {\mathbb{E}_{\theta}\bigl[\sum_{(i,j)}A_{ij}\bigr]},\\[4pt]
\text{mFDR}_{\theta}(\varphi) &=
\frac{\mathbb{E}_{\theta}\bigl[\sum_{(i,j)}(1-A_{ij})\varphi_{ij}(X)\bigr]}
     {\mathbb{E}_{\theta}\bigl[\sum_{(i,j)}\varphi_{ij}(X)\bigr]}.
\end{align} 
The mFDR is easier to estimate because it involves the ratio of expectations as opposed to the expectation of a ratio; for large numbers of tests it is asymptotically equivalent to the FDR
\citep{genovese2002operating}.


Given a target level $\alpha$, we seek a rule that maximizes TDR subject to
$\text{mFDR}_{\theta}\le\alpha$.
Define the \emph{structured $\ell$-value}
\[
\ell(x_{ij};Z_1,Z_2,\theta)=P\bigl(A_{ij}=0\mid X,Z_1,Z_2;\theta\bigr)
 = \ell\bigl(x_{ij},Z_{i,1},Z_{j,2};\theta\bigr),
\]
where, for block $(q,\ell)$,
\[
\ell(x_{ij},q,\ell;\theta)=
\frac{(1-\pi_{q\ell})\,g_0(x_{ij};\nu_0)}
     {\pi_{q\ell}\,g(x_{ij};\nu_{q\ell}) + (1-\pi_{q\ell})\,g_0(x_{ij};\nu_0)}.
\]
Unlike a classical BH $p$-value, the $\ell$-value pools information across \emph{all} observations via the shared parameters and cluster memberships, leading to substantial power gains.


In practice $(Z_1,Z_2,\theta)$ are unknown. Let $(\widehat Z_1,\widehat Z_2,\widehat\theta)$ denote their estimates obtained using the algorithm discussed in Section~\ref{sec:estimation}.  The hypothesis $H_{0,ij}$ is rejected whenever
\[
\ell \bigl(x_{ij};\widehat Z_1,\widehat Z_2,\widehat\theta\bigr)\le\tau,
\]
where $\tau=\tau(\alpha)$ is the largest threshold for which the plug-in estimate of mFDR does not exceed $\alpha$.

\section{Parameter estimation}\label{sec:estimation}

Estimating the model parameters  
\(\theta=\!(\balpha_1,\balpha_2,\Pi,\nu_0,\nu)
\)  
from the observed data matrix \(X=(x_{ij})\) is challenging, because the likelihood must be integrated over the unobserved adjacency matrix \(A\) and the block‐membership vectors \((Z_1,Z_2)\). In addition, the conditional distribution of the latent variables (\(A,Z_1,Z_2\)) given observed data does not have a closed-form expression. We therefore maximize a variational lower bound on the marginal likelihood via a \emph{variational expectation–maximization} (VEM) algorithm.

\subsection{Variational EM algorithm}\label{sec:vem}

For any distribution \(Q\) on \((A,Z_1,Z_2)\), the observed data log-likelihood can be decomposed as 
\begin{equation}\label{e:kl}
\log L(X;\theta)=
\E_Q\bigl[\log L(X,A,Z_1,Z_2;\theta)\bigr]
\;+\;H(Q)
\;+\;\mathrm{KL}\bigl(Q\,\Vert\,P_{A,Z_1,Z_2\mid X;\theta}\bigr),
\end{equation}
where \(H(Q)\) is the entropy and
\(\mathrm{KL}(\cdot\Vert\cdot)\) denotes Kullback–Leibler divergence.
Because the KL term is non-negative, the first two
terms form an \emph{evidence lower bound} (ELBO) on \(\log L(X;\theta)\).

The exact posterior
\(P_{A,Z_1,Z_2\mid X;\theta}\) is intractable due to the complex structure of our model, so we restrict \(Q\) to the
factorized family
\[
Q_{\beta_1,\beta_2}(A,Z_1,Z_2)\;=\;
P(A\mid Z_1,Z_2,X;\theta)
\prod_{i=1}^{n_1}\!\beta_{i,Z_{i,1},1}
\prod_{j=1}^{n_2}\!\beta_{j,Z_{j,2},2},
\]
with variational parameters
\(\beta_r=(\beta_{i,q,r})\in[0,1]^{n_r\times B_r}\)
satisfying \(\sum_{q=1}^{B_r}\beta_{i,q,r}=1\) for $r=1,2$.

\paragraph{E-step (update of \(\beta_1,\beta_2\)).}
Keeping \(\theta\) fixed at its current estimate
\(\theta^{(t)}\), we choose
\((\beta_1^{(t+1)},\beta_2^{(t+1)})\)
to maximize the ELBO.
This is equivalent to minimizing the KL divergence in \eqref{e:kl}
and yields the fixed-point updates (see Appendix)
\[
\beta_{i,q,1}\;\propto\;
\alpha_{q,1}\,
\exp\Bigl\{\!
    \sum_{j=1}^{n_2}\sum_{\ell=1}^{B_2}
    \beta_{j,\ell,2}\,d_{ij}^{q\ell}
\Bigr\},\qquad
\beta_{j,\ell,2}\;\propto\;
\alpha_{\ell,2}\,
\exp\Bigl\{\!
    \sum_{i=1}^{n_1}\sum_{q=1}^{B_1}
    \beta_{i,q,1}\,d_{ij}^{q\ell}
\Bigr\},
\]
where
\[
d_{ij}^{q\ell}\;=\;
\rho_{ij}^{q\ell}\,
\log\!\frac{\pi_{q\ell}\,g_{\nu_{q\ell}}(x_{ij})}{\rho_{ij}^{q\ell}}
+(1-\rho_{ij}^{q\ell})\,
\log\!\frac{(1-\pi_{q\ell})\,g_{\nu_0}(x_{ij})}{1-\rho_{ij}^{q\ell}}
\]
and
\(\rho_{ij}^{q\ell}=P(A_{ij}=1\mid Z_{i,1}=q,Z_{j,2}=\ell,X;\theta^{(t)})\).
In practice, 3–5 inner iterations suffice for convergence of the fixed-point updates.

\paragraph{M-step (update of \(\theta\)).} Assume for the moment the null and alternative densities are Gaussian. With \((\beta_1,\beta_2)\) fixed, maximizing the ELBO
gives closed-form updates:
\[
\hat\alpha_{q,1}=
\frac1{n_1}\sum_{i=1}^{n_1}\beta_{i,q,1},\qquad
\hat\alpha_{\ell,2}=
\frac1{n_2}\sum_{j=1}^{n_2}\beta_{j,\ell,2}, \qquad
\hat\pi_{q\ell}=
\frac{\sum_{i,j}\beta_{i,q,1}\beta_{j,\ell,2}\,\rho_{ij}^{q\ell}}
     {\sum_{i,j}\beta_{i,q,1}\beta_{j,\ell,2}},
\]
\[
\hat\nu_0:\;
\sigma_0^{2}=
\frac{\sum_{i,j}\!x_{ij}^{2}
      \sum_{q,\ell}\beta_{i,q,1}\beta_{j,\ell,2}(1-\rho_{ij}^{q\ell})}
     {\sum_{i,j}\sum_{q,\ell}
      \beta_{i,q,1}\beta_{j,\ell,2}(1-\rho_{ij}^{q\ell})},
\]
\[
\hat\nu_{q\ell}:\;
\mu_{q\ell}=
\frac{\sum_{i,j}\beta_{i,q,1}\beta_{j,\ell,2}\rho_{ij}^{q\ell}x_{ij}}
     {\sum_{i,j}\beta_{i,q,1}\beta_{j,\ell,2}\rho_{ij}^{q\ell}},\quad
\sigma_{q\ell}^{2}=
\frac{\sum_{i,j}\beta_{i,q,1}\beta_{j,\ell,2}\rho_{ij}^{q\ell}(x_{ij}-\mu_{q\ell})^{2}}
     {\sum_{i,j}\beta_{i,q,1}\beta_{j,\ell,2}\rho_{ij}^{q\ell}}.
\]
Updates for other tractable null and alternative densities, e.g., Gamma, can be derived similarly. 

The E- and M-steps are alternated until the algorithm converges.
Finally, the estimated memberships
\(\widehat Z_r=\arg\max_q\beta_{i,q,r} \ (r=1,2)\)  provide an interpretable biclustering of microbes and metabolites.

\paragraph{Initialization} We initialize parameters in the VEM algorithm in the following order: \(\beta_r, \rho, \Pi, \nu_0, \nu\) and \(\rho\) again before initiating the M-step of the algorithm. To initialize \(\beta_r\), we use k-means clustering on the rows and columns. We threshold the observed statistics at \(p\)-value 0.5 to obtain a rough estimate of \(\rho=(\rho_{ij}^{ql})\) so that we can initialize \(\Pi\). The parameters \(\nu_0\) and \(\nu\) are derived conditional on the initial cluster memberships. Finally, we improve the estimate of \(\rho\) conditional on the initial cluster memberships, \(\nu_0\) and \(\nu\). 

\subsection{Selecting the number of clusters}\label{sec:icl}

The VEM algorithm requires knowing the row and column block counts
\((B_1,B_2)\).
We choose them by maximizing the \emph{integrated classification
likelihood (ICL)}:
\[
\operatorname{ICL}(B_1,B_2)=
\E_{\widehat Q}\bigl[\log L(X,A,Z_1,Z_2;\hat\theta)\bigr]
-\operatorname{pen}_{\mathrm{BIC}}(B_1,B_2),
\]
where \(\widehat Q\) is the fitted variational distribution for the
given pair \((B_1,B_2)\),
\(\hat\theta\) the associated parameter estimate,
and
\[
\operatorname{pen}_{\mathrm{BIC}}(B_1,B_2)=
(B_1-1)\log n_1+(B_2-1)\log n_2
+\bigl[d_0+(1+d_1)B_1B_2\bigr]\log(n_1n_2)
\]
is the usual BIC penalty with \(d_0\) (null) and \(d_1\) (alternative) density parameters. The first two terms in the BIC penalty penalizes the number of parameters in $\alpha_r$, while the last term penalizes the number of parameters in $\Pi, \nu_0$ and $\nu$. By Equation \eqref{e:kl}, the ICL criterion can also be written in terms of the observed data likelihood, the entropy of \(\widehat Q\), and the BIC penalty. The entropy term favors
compact, well-separated clusters and therefore guards against
over-fitting.

\section{Numerical Experiments}\label{sec:experiments}

We first compare the proposed method to noisySBM \cite{rebafka2022powerful} to demonstrate the strength and necessity of the bipartite formulation. We then compare the proposed method to existing FDR control methods. 

\subsection{Comparison to noisySBM}

We sampled a bipartite network $A$ of size $n_1=40$ and $n_2 = 60$ with 2 row clusters and 3 column clusters. Nodes were assigned equal group probabilities among the rows and columns. The connectivity parameters were set such that $\pi_{1,1}=\pi_{2,2}=0.8$ and 0.1 elsewhere. Given $A$, observed data $x_{i,j}$ was sampled independently from a standard normal distribution if $A_{i,j}=0$ and from $\Nsc(\mu,0.25)$ if $A_{i,j}=1$. The matrix $X$ was used as input to the new procedure. The input to noisySBM was $\Xtilde = \begin{pmatrix}
    X_0 & X \\
    X' & X_1
\end{pmatrix}$, where entries in $X_0$ and $X_1$ were sampled independently from a standard normal distribution. This ensures that the latent graph $\Atilde$ given $\Xtilde$ corresponds to the same bipartite network $A$. 

We implemented both the new procedure and noisySBM under the ideal scenario that the null density and the true number of clusters are both known. Figure \ref{fig:biSBMvsSBM} compares their empirical FDR, TDR, and run time of noisySBM across different values of $\mu$.  At $\mu=1$, noisySBM shows severely inflated FDR and a slightly lower TDR than the new method. This is because noisySBM produces poor estimation of latent block memberships which compromises its performance on hypothesis testing. The new procedure, while clustering rows imperfectly, achieves better column clustering and therefore a smaller FDR. As the signal increases, the new method’s FDR approaches the nominal level, whereas noisySBM’s FDR first falls and then rises because it increasingly misclassifies edges between nodes of the same type.
TDRs are nearly identical for both methods across most settings. 
Computationally, however, the new procedure is at least 20 times faster: it solves two smaller problems, whereas noisySBM must infer from the data that cross-type connections are absent---a task that markedly slows its inference. Similar advantages of the bipartite SBM formulation---in both speed and accuracy of community detection---have also been reported when the observation is noiseless, i.e., an adjacency matrix \cite{larremore2014efficiently}.
\begin{figure}
    \centering
    \includegraphics[width=\linewidth]{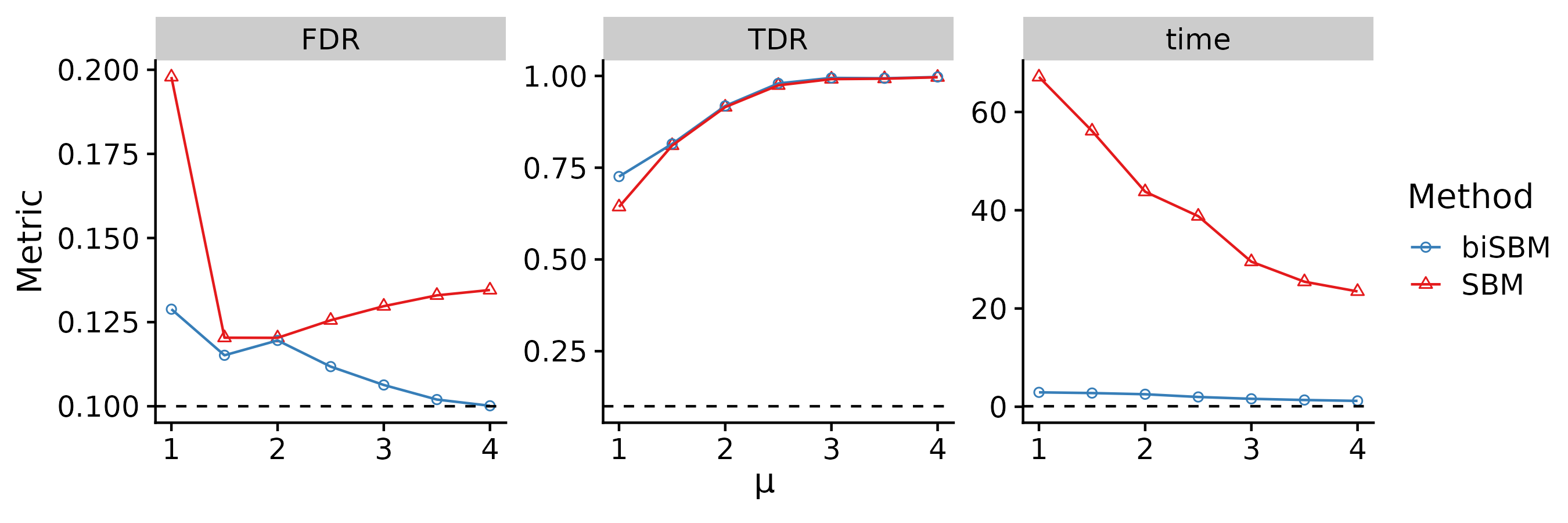}
    \caption{Performance of noisySBM versus its bipartite formulation in multiple testing, evaluated as a function of the alternative mean. Dashed line indicates the nominal level 0.1.}
    \label{fig:biSBMvsSBM}
\end{figure}

\subsection{Comparison to other FDR control methods}

We also compared the proposed new procedure with the BH procedure \cite{benjamini1995controlling}, Storey's $q$-value procedure \cite{storey2004strong}, and the Sun and Cai procedure (SC) \cite{sun2009large}. We implemented the new procedure with known null density and used the ICL criterion to perform model selection. For fair comparison, we also implemented the SC procedure with known null density.

We simulated data in the following scenarios. In all scenarios, undirected bipartite graphs of dimension $n_1=150$ by $n_2=200$ were generated and the standard normal distribution was chosen as the null.  
\begin{enumerate}
    \item[(a)] Data  were sampled with $B_1 = B_2 = 3$ latent biclusters and equal group probabilities $\alpha_{q,1}=\alpha_{q,2} = 1/3$ for $q=1,2,3$. The connectivity parameters were set such that $\pi_{q,l} = 0.1 \cdot \bone(q\ne l) + 0.8 \cdot \bone(q=l)$ for $q,l = 1,2,3$ where $\bone(\cdot)$ is the indicator function. For the alternative distributions, we set Gaussian means $\mu_{1,1}=\mu_{2,2}=\mu_{3,3}=1$ and $\mu_{q,l}=3$ elsewhere so that sparser clusters have a larger alternative mean.  
\end{enumerate}
Besides the above modular bipartite graph, we also considered fixed graphs with other topological properties. Given $A$, data were sampled independently from $N(0,1)$ is $A_{i,j}=0$ and from $N(2,1)$ if $A_{i,j} = 1$. 
\begin{enumerate}
    \item[(b)] The latent graph $A$ is a fully nested bipartite graph (Figure \ref{fig:biSBMnetwork}B) \cite{pavlopoulos2018bipartite}. In other words, there is a single `generalist' in each vertex type that is connected with all the nodes in the other vertex type. 
    \item[(c)] The latent graph $A$ is generated according to a bipartite preferential attachment model (Figure \ref{fig:biSBMnetwork}C) \cite{guillaume2006bipartite}. At each step, a new type I vertex is added and its degree $d$ is sampled uniformly at random from the set $\{2,3,4,5,6\}$. Then, for each of the $d$ edges of the new vertex, either a new type II vertex is added (with probability  $1-\lambda$) or one is picked among the preexisting ones using preferential attachment (with probability $\lambda$). We set the parameter $\lambda=0.8$, which is the average ratio of preexisting type II vertices to which a new type I vertex is connected. 
\end{enumerate}
Figure \ref{fig:biSBMnetwork} presents illustrations of the latent bipartite network in each scenario.

\begin{figure}
    \centering
    \includegraphics[width=\linewidth]{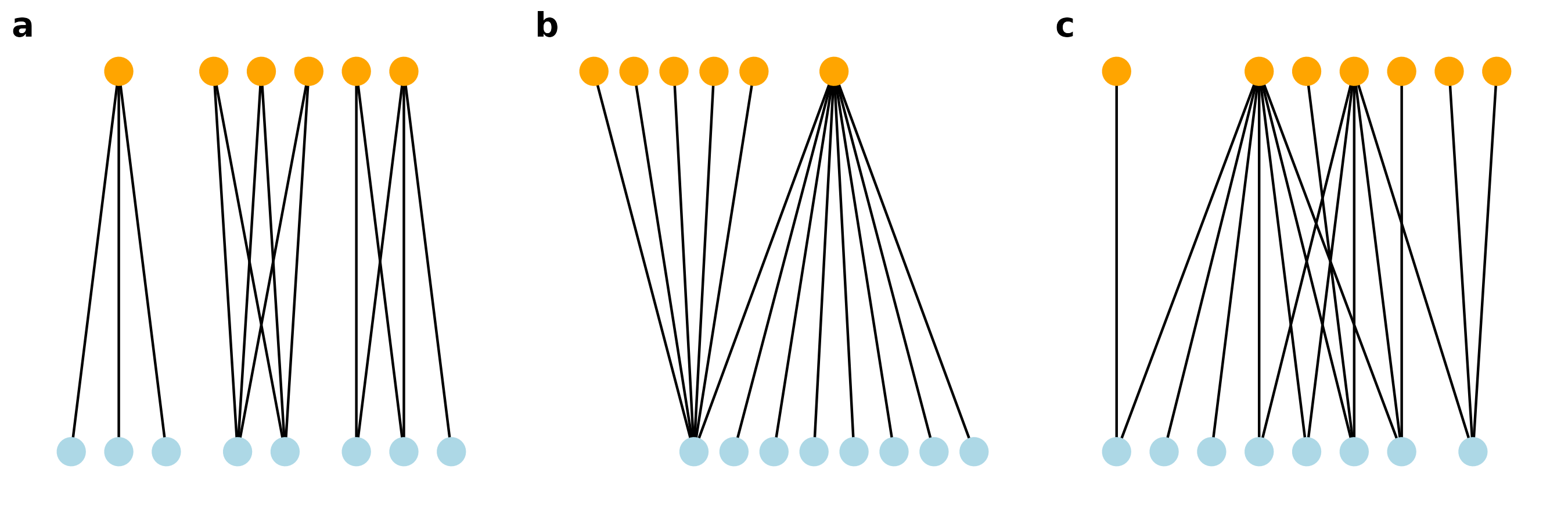}
    \caption{Illustrations of the latent bipartite network in the three scenarios: (a) three biclusters, (b) nested, and (c) preferential attachment. Color of the vertices indicates types of vertices in the graph.} 
    \label{fig:biSBMnetwork}
\end{figure}

We simulated 100 data sets for each scenario and applied every testing procedure at nominal FDR level $\alpha\in \{0.005,0.025, 0.05, 0.1, 0.15, 0.25\}$. Figure \ref{fig:biSBMOthers} provides the ROC curves of the empirical FDR vs TDR as functions of $\alpha$. 

Across all scenarios, the BH and Storey's $q$-value procedures behave almost identically: they keep FDR near the nominal levels but achieve only modest TDR, with Storey’s method being slightly less conservative. The SC procedure controls the FDR at nominal levels in scenarios (a) and (c) but inflates FDR in scenario (b); its power is only marginally better than the $q$-value approach. 

By contrast, the new procedure consistently outperforms existing methods across all scenarios. In scenario (a), its FDR stays close to the nominal level while its TDR is much higher. In 90\% of the simulations the ICL criterion correctly selects three biclusters, and this accurate recovery of the latent structure markedly boosts the new method's power. In scenario (b), although the data do not follow the assumed model, the ICL criterion chooses two biclusters in 83\% of the simulations---one ``generalist'' cluster in each vertex set and a complementary ``specialist'' cluster. Estimated connection probabilities reflect the nested structure ($\approx 0.99$ for generalist–specialist edges, $\approx 0$ for specialist–specialist edges), enabling FDR below nominal levels and high TDR---something competing methods miss without explicit modeling. In scenario (c), the new method shows a slight FDR inflation, yet its ROC curve still dominates those of all other procedures, indicating robustness to model misspecification. In most simulations, the ICL criterion discovers three clusters in type-I vertices (high-, medium-, and low-degree nodes) and one in type-II vertices, yielding a meaningful partition and sustained power despite imperfect model fit.

Overall, by learning an appropriate biclustering, the new procedure maintains or improves FDR control and delivers substantially higher power, while existing methods remain conservative but underpowered.

\begin{figure}
    \centering
    \includegraphics[width=\linewidth]{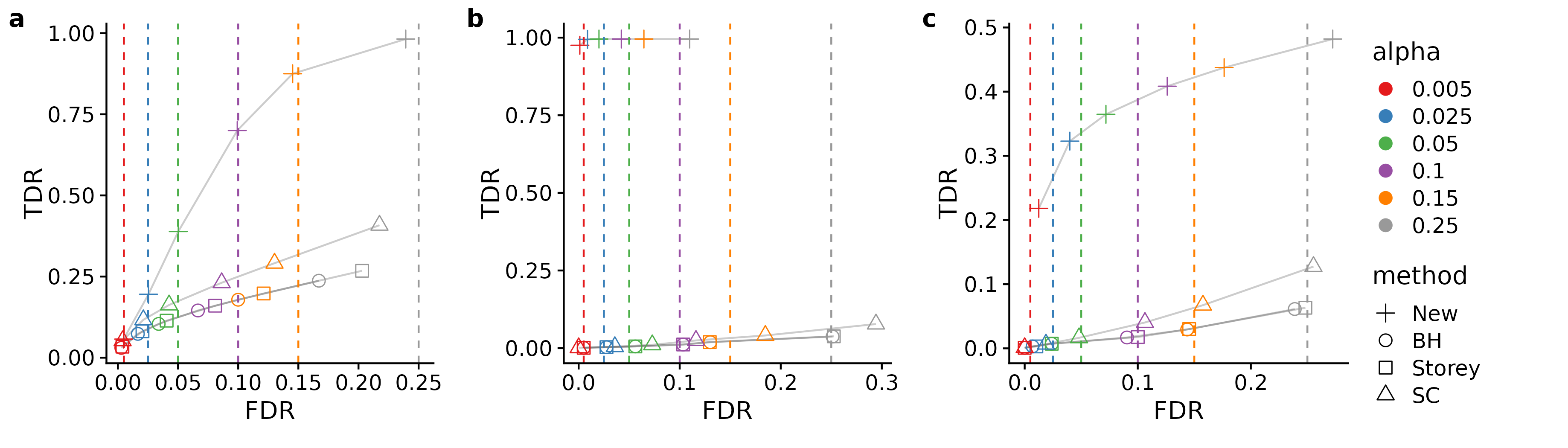}
    \caption{Plot of the empirical $({\rm FDR}, {\rm TDR})$ as functions of the nominal level $\alpha$ for the new procedure, BH, Storey's $q$-value, and the SC procedure. Dashed lines indicate the nominal level $\alpha$.}
    \label{fig:biSBMOthers}
\end{figure}

\section{Application to Bacterial Vaginosis}\label{sec:app}

\subsection{Data description and preprocessing}


We analyzed paired microbiome–metabolome data from 131 Rwandan women reported in \citep{mcmillan2015multi}. 16S rRNA sequencing yielded 51 bacterial taxa after the authors’ initial quality filters.  We removed one taxon present in only 13 participants and merged duplicate \textit{Bacteroides} counts, leaving 49 distinct taxa.  To correct for varying sequencing depth, we applied the modified centred log-ratio (mCLR) transform, which rescales positive counts while leaving zeros unchanged \citep{ma2020joint}. Gas-chromatography mass spectrometry profiled 128 vaginal metabolites; log-transformed abundances were provided in the original dataset. Among the 131 participants, 79 were classified as “normal”, 23 as BV, 22 as “intermediate” and 7 had no clinical diagnosis.  Following \citep{mcmillan2015multi}, we pooled the BV and intermediate groups into a single disease cohort (\(n=45\)) and compared it with the normal cohort (\(n=79\)) to identify metabolic alterations associated with BV.

\subsection{Test statistics}\label{sec:teststat}

The proposed inference procedure requires a matrix \(X=(x_{ij})\) of approximately standard normal \(z\)-scores, each summarizing the association between the \(i\)-th microbe and the \(j\)-th metabolite.  Let the two paired data blocks be
\(Y_1\in\mathbb{R}^{m\times n_1}\) (microbiome) and
\(Y_2\in\mathbb{R}^{m\times n_2}\) (metabolome), where the \(k\)-th row
corresponds to the same individual in both matrices.

Write \(\hat\mu_{1}\) and \(\hat\mu_{2}\) for the column means of \(Y_1\) and
\(Y_2\), and let \(\hat\sigma^{2}_{1,i}\) and \(\hat\sigma^{2}_{2,j}\) denote the
corresponding sample variances for each feature.  After
standardizing each column,
\[
\Ytilde_{1,k,i}=\frac{Y_{1,k,i}-\hat\mu_{1,i}}{\sqrt{\hat\sigma^{2}_{1,i}}},
\qquad
\Ytilde_{2,k,j}=\frac{Y_{2,k,j}-\hat\mu_{2,j}}{\sqrt{\hat\sigma^{2}_{2,j}}},
\]
the empirical Pearson correlation is
\(
\rhohat_{ij}=m^{-1}\sum_{k=1}^{m}\Ytilde_{1,k,i}\Ytilde_{2,k,j}.
\)
Following \cite{cai2016large}, an asymptotically normal test statistic is
\[
x_{ij}=\frac{2\rhohat_{ij}}{\sqrt{s_{ij}/m}}
\;\xrightarrow{d}\;
N(0,1),
\]
where
\(
s_{ij}=m^{-1}\sum_{k=1}^{m}
\bigl(2\Ytilde_{1,k,i}\Ytilde_{2,k,j}-\rhohat_{ij}\Ytilde_{1,k,i}
-\rhohat_{ij}\Ytilde_{2,k,j}\bigr)^{2}
\).

To detect changes in microbe–-metabolite correlation between two clinical
subgroups (\(d=1,2\)) with sample sizes \(m_d\),
let \(\rhohat^{(d)}_{ij}\) and \(s^{(d)}_{ij}\) be the
within-group estimates defined above.  The null hypothesis
\(H_{0,ij}:\rho^{(1)}_{ij}=\rho^{(2)}_{ij}\) is tested by
\begin{equation}\label{eq:twosample}
    x_{ij}=\frac{2\bigl(\rhohat^{(1)}_{ij}-\rhohat^{(2)}_{ij}\bigr)}
           {\sqrt{s^{(1)}_{ij}/m_1+s^{(2)}_{ij}/m_2}}
\;\xrightarrow{d}\;
N(0,1).
\end{equation}
The resulting \(z\)-value matrix \(X\) serves as the input to the bipartite graph-based multiple-testing procedure described in Section \ref{sec:method}.

\subsection{Results}

Figure~\ref{fig:BVrest}A shows the distribution of the observed $z$-scores, defined in \eqref{eq:twosample}. Although the test statistics appear to be normally distributed, they deviate noticeably from the standard normal curve; the marginal density recovered by our method offers a better fit than the standard normal. Panel B demonstrates that, across every significance threshold, the new procedure yields more discoveries than competing methods, confirming its superior power. Panels C and D display heatmaps of the observed $z$-scores and the network inferred by our method at $\alpha$=0.1\%. Rows and columns are ordered by the clusters identified by the algorithm. In Panel D, each block is colored by its alternative mean values, indicating up-regulation (negative mean) and down-regulation (positive mean) in BV. Two blocks with the largest mean in absolute value are highlighted. The most up-regulated block contains the genera {\it Leptotrichia, Leptotrichiaceae\_unclassified}, and {\it Sneathia}---all emerging BV pathogens \cite{thilesen2007leptotrichia,theis2021sneathia, laniewski2021bacterial}. The most down-regulated block comprises {\it Lactobacillus, Lactobacillus\_crispatus, Lactobacillus\_gasseri/johnsonii, Anaerococcus, Corynebacterium, Finegoldia} and {\it Escherichia/Shigella}; the first three taxa are known to be less abundant in BV patients. 
Both groups of taxa are linked to the same set of metabolites---{\it 2-hydroxyisovalerate, succinate} and three uncharacterized metabolites. {\it 2-hydroxyisovalerate} has already been associated with high diversity and clinical BV \cite{mcmillan2015multi}. Although {\it succinate} was not significant in that study's univariate tests, our multivariate approach highlights its role through ``guilt by association'', consistent with other reports implicating {\it succinate} in BV \cite{srinivasan2015metabolic}. Association between Lactobacillus species and three uncharacterized metabolites may shed light on the function of these metabolites. Together, these results reveal broad shifts in microbial metabolism during BV and pinpoint specific taxa–-metabolite links for future experimental validation.

\begin{figure}
    \centering
    \includegraphics[width=\linewidth]{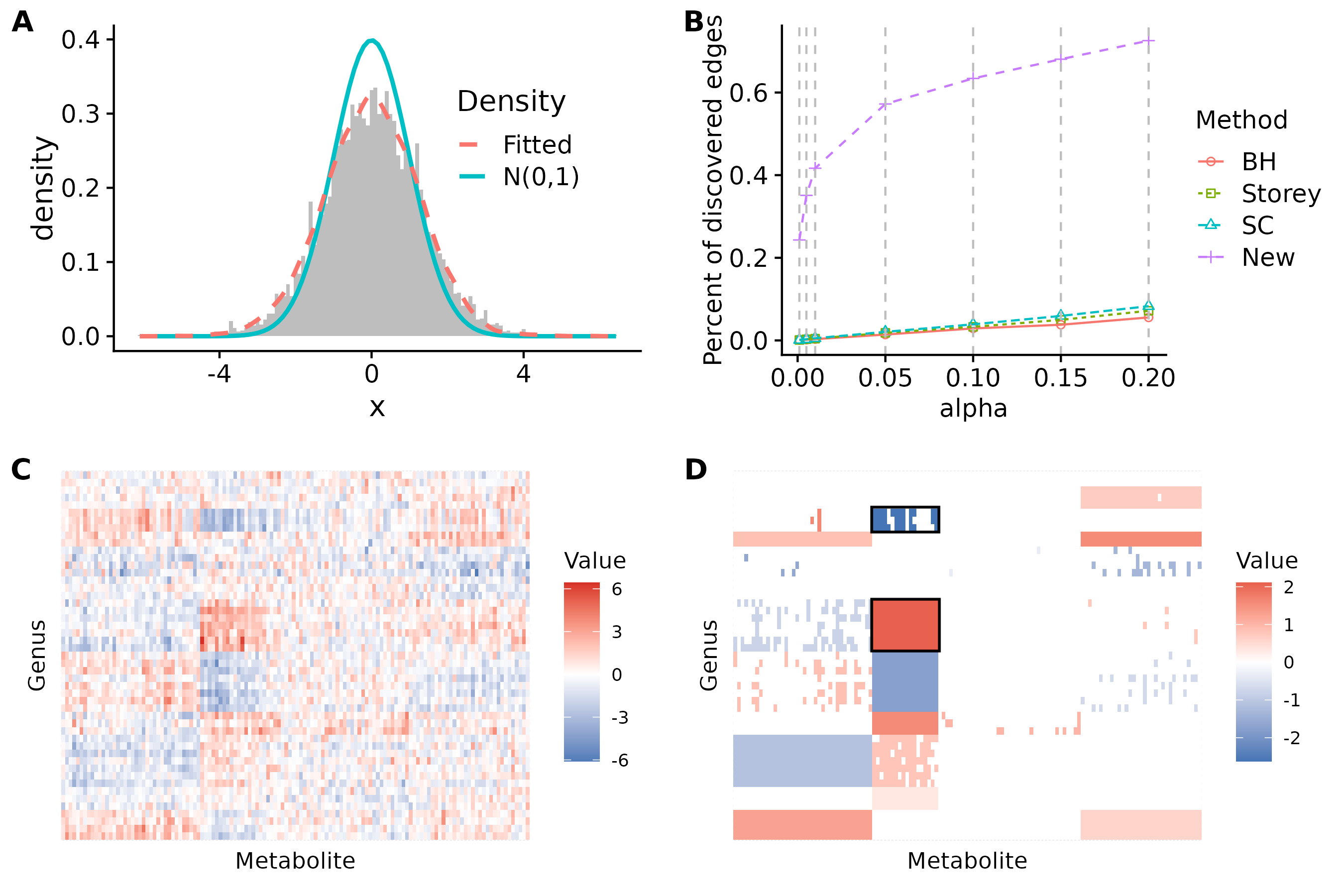}
    \caption{(A) Histogram of observed z-scores compared to the standard normal distribution and to the estimated marginal distribution by the proposed approach. (B) Percent of rejected edges as a function of the significance level $\alpha$ for the different procedures. (C) Heatmap of the observed z-scores ordered by the inferred clustering. (D) Heatmap of the inferred network at $\alpha=0.1\%$ with colors indicating the estimated mean values in each block. Two blocks with the largest mean in absolute value are highlighted.}
    \label{fig:BVrest}
\end{figure}

\section{Discussion}\label{sec:diss}


Joint analysis of microbiome and metabolomic data can illuminate the metabolic circuitry that underlies host–microbe interactions and, ultimately, guide the design of microbiome-based therapeutics. This paper introduced a \emph{bipartite noisy stochastic block model} that models the observed $z$-score matrix with a latent microbe–-metabolite interaction graph.  
By exploiting the latent block structure, the method (i) achieves reliable FDR control, (ii) delivers substantially higher power than existing
procedures, and (iii) provides a biologically interpretable biclustering of taxa and metabolites.  
Simulations confirmed the gains in speed and accuracy over the original (noisy) SBM and over conventional FDR tools. The BV case study further demonstrated how the inferred modules translate into concrete, experiment-ready hypotheses on altered metabolic activity.


In our model, we can view the indicator matrix $A$ as an unobserved parameter drawn from a structured prior---in our case, a bipartite SBM.  Estimating the hyper-parameters via marginal maximum likelihood, and replacing them in the $\ell$-value formula, is analogous to the empirical Bayes framework originally introduced in \citep{efron2001empirical}.  The key difference is that we encode \emph{dependence} through the latent graph, so the procedure can borrow strength without assuming weak $p$-value correlations.


Like all model-based methods, our approach relies on reasonable hyper-parameter estimates.  If the signal-to-noise ratio (SNR) is extremely low, estimation may become unstable, leading to FDR inflation. In our model, SNR depends on both the magnitude of the alternative mean and the density of the latent graph. Weak but dense signals can still be well estimated; weak and sparse signals may require larger sample sizes or an informative prior on $\Pi$. 

Our method assumes each $x_{ij}$ has a tractable null density.
Pearson- or Spearman-based $z$-scores satisfy this requirement
\citep{cai2016large}.  Detecting non-linear dependence would call for statistics such as distance correlation \cite{szekely2007measuring}; their null distribution is complex and often approximated, a step we leave for future work.

The variational EM (VEM) algorithm used by our method scales linearly in the number of edges and converges rapidly, yet the search over $(B_1,B_2)$ via the ICL criterion can still be time-consuming.  A promising alternative is the greedy ICL maximization of \citep{kilian2024enhancing}, which iteratively reassigns vertices and naturally prunes empty blocks.


This work can be extended in several directions. First, if the microbiome and metabolomic data are subject to common confounding, such as batch effects, confounder-adjusted partial correlations could be plugged into the model exactly as ordinary correlations, provided their null distribution is known or well estimated. Second, longitudinal microbiome studies are proliferating. Extending the current model to a dynamic bipartite SBM could track the birth, death, and rewiring of modules over time. Lastly, the bipartite framework naturally generalizes to a multipartite noisySBM, allowing simultaneous testing across, e.g., microbes, metabolites, and transcripts. 

\section{Appendix}

\subsection{The variational E-step}

In the variational E-step, we need to maximize the ELBO
\[
\bbE_Q [ \log L(X, A, Z_1, Z_2; \theta) ]  - \bbE_Q[\log Q]
\]
to solve for the variational parameters $\beta_1$ and $\beta_2$.

Denote by $\Asc=\{(i,j): 1\le i \le n_1, 1\le j \le n_2\}$ the set of all possible pairs that we wish to test. The complete-data likelihood can be rewritten as 
\begin{align*}
    L(X, A, Z_1, Z_2; \theta) = & ~ \prod_{\underset{A_{i,j}=0}{(i,j)\in \Asc:}} g_{\nu_0}(x_{i,j}) \prod_{q=1}^{B_1}\prod_{l=1}^{B_2} \prod_{\underset{Z_{i,1}=q, Z_{j,2} = l}{(i,j): A_{i,j}=1 }}g_{\nu_{q,l}}(x_{i,j}) \times \\
    & \prod_{q=1}^{B_1}\prod_{l=1}^{B_2} \pi_{q,l}^{M_{q,l}} (1-\pi_{q,l})^{\Mtilde_{q,l}} \times \prod_{r=1}^2 \prod_{q=1}^{B_r} \alpha_{q,r}^{\sum_{i=1}^{n_r}\bone(Z_{i,r}=q)},
\end{align*}
where 
\begin{align*}
M_{q,l} & = \# \{(i,j)\in \Asc: A_{i,j} = 1, Z_{i,1}=q, Z_{j,2}=l \},\\
\Mtilde_{q,l}& = \# \{(i,j)\in \Asc: A_{i,j} = 0, Z_{i,1}=q, Z_{j,2}=l \}.
\end{align*}
Let $\Etilde_{\beta_1,\beta_2}$ and $\Ptilde_{\beta_1,\beta_2}$ denote, respectively, the expected value and probability distribution function with respect to the variational distribution. Substituting $Q$ with the factorized variational distribution, we get 
\begin{align*}
    \Etilde_{\beta_1,\beta_2}  [ \log L(X, A, Z_1, Z_2; \theta) ] 
    & = \sum_{(i,j)\in \Asc} {\Ptilde}_{\beta_1,\beta_2}(A_{i,j}=0)\log g_{0,\nu_0}(x_{i,j})  \\ & +\sum_{q=1}^{B_1}\sum_{l=1}^{B_2} \sum_{{(i,j)\in \Asc }} \Ptilde_{\beta_1,\beta_2}(A_{i,j}=1, Z_{i,1}=q, Z_{j,2} = l) \log g_{\nu_{q,l}}(x_{i,j})  \\
    &+ \sum_{q=1}^{B_1}\sum_{l=1}^{B_2} \sum_{{(i,j)\in \Asc }} \Ptilde_{\beta_1,\beta_2}(A_{i,j}=1, Z_{i,1}=q, Z_{j,2} = l) \log \pi_{q,l}  \\
    & +\sum_{q=1}^{B_1}\sum_{l=1}^{B_2}\sum_{{(i,j)\in \Asc }} \Ptilde_{\beta_1,\beta_2}(A_{i,j}=0, Z_{i,1}=q, Z_{j,2} = l) \log (1-\pi_{q,l} )  \\
    & +\sum_{r=1}^2 \sum_{q=1}^{B_r} \sum_{i=1}^{n_r}\Etilde_{\beta_1,\beta_2}[\bone(Z_{i,r}=q)] \log \alpha_{q,r}
\end{align*}
To calculate the entropy of the factorized variational distribution $Q$, note that
\begin{align*}
   \Etilde_{\beta_1,\beta_2}  [\log Q ] & =  \sum_{q=1}^{B_1}\sum_{l=1}^{B_2} \sum_{{(i,j)\in \Asc }} \Ptilde_{\beta_1,\beta_2}(A_{i,j}=1, Z_{i,1}=q, Z_{j,2} = l) \log \rho_{ij}^{ql}  + \\
    & \quad  \sum_{q=1}^{B_1}\sum_{l=1}^{B_2}\sum_{{(i,j)\in \Asc }} \Ptilde_{\beta_1,\beta_2}(A_{i,j}=0, Z_{i,1}=q, Z_{j,2} = l) \log (1-\rho_{ij}^{ql})  .
\end{align*}
And 
\[
\Etilde_{\beta_1,\beta_2} [\log (\prod_{i=1}^{n_1} \beta_{i,Z_{i,1},1} )] = \sum_{q=1}^{B_1} \sum_{i=1}^{n_1}\Etilde_{\beta_1,\beta_2}[\bone(Z_{i,r}=q)] \log \beta_{i,q,1}.
\]
By Bayes rule, the conditional distribution of $A$ given block memberships $Z_1$ and $Z_2$ follows a Bernoulli distribution with conditional probability 
\[
P(A_{i,j} = 1\mid Z_{i,1}=q, Z_{j,2}=l, X) = \frac{\pi_{q,l} g_{\nu_{q,l}}(x_{i,j}) }{\pi_{q,l} g_{\nu_{q,l}}(x_{i,j})  + (1-\pi_{q,l}) g_{\nu_0}(x_{i,j}) } := \rho^{ql}_{ij}.
\]
Putting all the pieces together, we obtain 
\begin{align*}
\mbox{ELBO} =  - &  \sum_{q=1}^{B_1}\sum_{l=1}^{B_2} \sum_{{(i,j)\in \Asc }} \beta_{i,q,1}\beta_{j,l,2} \left\{ \rho_{ij}^{ql} \log \rho_{ij}^{ql}+ (1-\rho_{ij}^{ql} ) \log (1-\rho_{ij}^{ql})\right\} +\\
    & \sum_{q=1}^{B_1}\sum_{l=1}^{B_2} \sum_{{(i,j)\in \Asc }} \beta_{i,q,1}\beta_{j,l,2} \left\{ \rho_{ij}^{ql} \log \pi_{q,l}+ (1-\rho_{ij}^{ql} ) \log (1-\pi_{q,l})\right\} + \\
    & \sum_{(i,j)\in \Asc} \log g_{0,\nu_0}(x_{i,j}) \sum_{q=1}^{B_1} \sum_{l=1}^{B_2} \beta_{i,q,1} \beta_{j,l,2} (1 - \rho_{ij}^{ql}) + \\
    & \sum_{q=1}^{B_1}\sum_{l=1}^{B_2} \sum_{{(i,j)\in \Asc }} \rho_{ij}^{ql} \beta_{i,q,1} \beta_{j,l,2}  \log g_{\nu_{q,l}}(x_{i,j}) + \\
    & \sum_{q=1}^{B_1}\sum_{i=1}^{n_1} \beta_{i,q,1} \log \frac{\alpha_{q,1}}{\beta_{i,q,1}} + \sum_{l=1}^{B_2} \sum_{j=1}^{n_2}\beta_{i,l,2} \log \frac{\alpha_{l,2}}{\beta_{j,l,2}}.
\end{align*}
After rearranging terms, we can rewrite the ELBO as
\begin{align*}
    \mbox{ELBO}  & = \sum_{q=1}^{B_1}\sum_{i=1}^{n_1} \beta_{i,q,1} \log \frac{\alpha_{q,1}}{\beta_{i,q,1}} + \sum_{l=1}^{B_2} \sum_{j=1}^{n_2}\beta_{i,l,2} \log \frac{\alpha_{l,2}}{\beta_{j,l,2}} + \sum_{q=1}^{B_1}\sum_{l=1}^{B_2} \sum_{{(i,j)\in \Asc }} \beta_{i,q,1}\beta_{j,l,2} d_{ij}^{ql}.
\end{align*}
The partial derivative of the ELBO with respect to $\beta_{i,q,r}$ is given by 
\begin{align*}
\frac{\partial }{\partial \beta_{i,q,1}} \mbox{ELBO}  & = \log \alpha_{q,1} - \log \beta_{i,q,1} - 1 + \sum_{l=1}^{B_2} \sum_{j=1}^{n_2} \beta_{j,l,2} d_{ij}^{ql},\\
\frac{\partial }{\partial \beta_{j,l,2}} \mbox{ELBO}  & = \log \alpha_{q,2} - \log \beta_{j,l,2} - 1 + \sum_{q=1}^{B_1} \sum_{i=1}^{n_1} \beta_{i,q,1} d_{ij}^{ql}.\end{align*}
Scenario the partial derivative to zero yields
\begin{equation}\label{eq:var}
    \beta_{i,q,1} = \alpha_{q,1} \exp\left(\sum_{l=1}^{B_2} \sum_{j=1}^{n_2} \beta_{j,l,2} d_{ij}^{ql} - 1\right),\quad \beta_{j,l,2} = \alpha_{l,2} \exp\left(\sum_{q=1}^{B_1} \sum_{i=1}^{n_1} \beta_{i,q,1} d_{ij}^{ql} - 1\right).
\end{equation}
The probabilities also satisfy the constraint $\sum_{q=1}^{B_r}\beta_{i,q,r} = 1$ for all $i$ and $r$. There are no explicit solutions to these equations \eqref{eq:var}, but they can be achieved via fixed-point iterations.

 \bibliographystyle{unsrt}
 \bibliography{365references}
\end{document}